\documentclass[aps,prd,twocolumn,superscriptaddress, nofootinbib,floats,floatfix]{revtex4-1}
\usepackage{bm}
\usepackage{amsmath}
\usepackage{amsfonts}
\usepackage{aas_macros}
\usepackage{graphicx}
\usepackage[hidelinks]{hyperref}
\usepackage{booktabs}
\usepackage{color}
\usepackage[dvipsnames]{xcolor}
\usepackage[capitalise]{cleveref}
\usepackage[caption=false]{subfig}

\newcommand\sun{\hbox{$\odot$}}
\newcommand\degr{\hbox{$^\circ$}}
\newcommand\arcmin{\hbox{$^\prime$}}

\creflabelformat{equation}{#2\textup{#1}#3}

\newcommand{\gcon}{2.1 \times 10^{-15}}

\begin{document}

\title{Constraints on equivalence principle violation from gamma ray bursts}

\author{D. J. Bartlett}
\email{deaglan.bartlett@physics.ox.ac.uk}
\affiliation{Astrophysics, University of Oxford, Denys Wilkinson Building, Keble Road, Oxford, OX1 3RH, UK}
\author{D. Bergsdal}
\affiliation{Particle and Astroparticle Physics, Department of Physics, KTH Royal Institute of Technology, SE 106 91, Stockholm, Sweden}
\affiliation{The Oskar Klein Centre, Department of Physics, Stockholm University, AlbaNova University Centre, SE 106 91 Stockholm, Sweden}
\author{H. Desmond}
\affiliation{Astrophysics, University of Oxford, Denys Wilkinson Building, Keble Road, Oxford, OX1 3RH, UK}
\author{P. G. Ferreira}
\affiliation{Astrophysics, University of Oxford, Denys Wilkinson Building, Keble Road, Oxford, OX1 3RH, UK}
\author{J. Jasche}
\affiliation{The Oskar Klein Centre, Department of Physics, Stockholm University, AlbaNova University Centre, SE 106 91 Stockholm, Sweden}

\begin{abstract}
    Theories of gravity that obey the Weak Equivalence Principle have the same Parametrised Post-Newtonian parameter $\gamma$ for all particles at all energies. The large Shapiro time delays of extragalactic sources allow us to put tight constraints on differences in $\gamma$ between photons of different frequencies from spectral lag data, since a non-zero $\Delta \gamma$ would result in a frequency-dependent arrival time. The majority of previous constraints have assumed that the Shapiro time delay is dominated by a few local massive objects, although this is a poor approximation for distant sources. In this work we consider the cosmological context of these sources by developing a source-by-source, Monte Carlo-based forward model for the Shapiro time delays by combining constrained realisations of the local density field using the Bayesian origin reconstruction from galaxies algorithm with unconstrained large-scale modes. Propagating uncertainties in the density field reconstruction and marginalising over an empirical model describing other contributions to the time delay, we use spectral lag data of Gamma Ray Bursts from the BATSE satellite to constrain $\Delta \gamma < \gcon$ at $1 \sigma$ confidence between photon energies of $25 {\rm \, keV}$ and $325 {\rm \, keV}$.
\end{abstract}

\maketitle

\section{Introduction}

A basic feature of metric theories of gravity is that objects on either timelike or null geodesics experience apparent time delays due to motion through regions of varying spacetime curvature, caused by the difference between proper and observer time induced by a gravitational field. This effect, first derived by Shapiro in 1964 \citep{Shapiro_1964}, has now been accurately measured in the Solar System: the best constraint on the fractional deviation of the time delay from the prediction of General Relativity (GR) is currently at the $10^{-5}$ level, using data from the radio link with the Cassini spacecraft \cite{Bertotti_2003}. In the Parametrised Post-Newtonian (PPN) framework the time delay is governed by the light-bending parameter $\gamma$ \cite{Will_1972}. For sources at greater distance, however, the lack of knowledge of the time of emission of a signal precludes direct measurement of the delay.

Nevertheless, the time delay effect has found use in testing various aspects of the theory of gravity and the standard model of cosmology. This is done by \textit{comparing} time delays, either between nearby geodesics or between different types of object following the same geodesic. The former may be achieved by comparing the time of reception of photons originating from a common source but traversing different paths due to gravitational lensing. When the source is a time-varying quasar and the lens a single massive elliptical galaxy or cluster, this method has been used to constrain the Hubble parameter, $H_0$, on which time delay distances depend \cite{Holicow_2020,Birrer_2020}.
Conversely, comparing time delays between different objects travelling along the same geodesic allows the Weak Equivalence Principle (WEP) to be tested by investigating whether they experience time delays identically. This can be achieved either when the source emits light at varying frequencies which can be independently measured \citep[][]{Luo_2016}, or when it emits other types of energy in addition to light, such as gravitational waves \cite{GW170817}. Within the PPN framework, this constrains the difference between their $\gamma$ factors. 

In general metric theories the time delay is proportional to the integral of the fluctuation in Newtonian potential $\phi$ along the line of sight to the source; within the PPN framework it has a prefactor $1+\gamma$ where $\gamma=1$ in GR. Predicting time delays is therefore equivalent to determining $\delta \phi$. In typical analyses, $\delta \phi \simeq \phi$ is modelled as arising from one or a few isolated sources near the line of sight that are believed to be predominantly responsible for sourcing the potential. However, the long range of the gravitational potential ($\phi \sim 1/r$) casts doubt on the multiple point masses approximation, since $\phi$ is sensitive to the distribution of distant sources and thus should be considered in a cosmological context \cite{Nusser_2016,Minazzoli_2019}.

The aim of this work is to account fully for the contributions to the time delay from all mass in the non-linear cosmological density field. As in \cite{Nusser_2016}, we consider an unconstrained contribution from distant sources, however we also combine this with the contribution from local structures using constrained density fields generated by the BORG algorithm \cite{BORG_1,BORG_2,BORG_3,BORG_4,Lavaux_2016} to produce a Monte Carlo-based source-by-source forward model for the expected Shapiro time delay.

We apply our method to high-energy astrophysical sources that have previously been used to test the Equivalence Principle for photons of varying wavelengths. Marginalising  over  uncertainties in the density field reconstruction and parameters describing non-Equivalence-Principle-violating contributions, we compare our predictions to the observations via  a  Markov  Chain  Monte  Carlo (MCMC) algorithm. We find our constraints are $\sim$40 times tighter than literature results, illustrating the
benefit of using
complete mass distributions when studying non-local relativistic effects such as time delays in a cosmological setting.

In \Cref{sec:theory} we discuss time delays in the context of metric theories of gravity, including the importance of long wavelength modes in calculating the Shapiro time delay.
We describe the Gamma Ray Burst (GRB) observations in \Cref{sec:Observational data} and detail our inference methods in \Cref{sec:Methods}. The results are presented in \Cref{sec:Results}.
A discussion of systematic uncertainties, future applications of our methods and a comparison to the literature are presented in \Cref{sec:Discussion}. We conclude in \Cref{sec:Conclusions}.

\section{Time delay differences in metric theories of gravity}
\label{sec:theory}

\subsection{Equivalence principle violation}

Consider a perturbed FRW-like metric in the Newtonian gauge,
\begin{equation}
	{\rm d} s^2 = - \left( 1 + 2 \delta\phi \right) {\rm d} t^2 + a^2 \left( t \right) \left( 1 - 2 \gamma \delta\phi \right) {\rm d} \bm{r}^2,
\end{equation}
where we have introduced the PPN parameter $\gamma$ to allow for deviations from General Relativity (where $\gamma=1$).

For a massless particle, the gravitational time delay to a source at distance $r_s$ is given by $\left( 1 + \gamma \right) t_{\rm grav}$, where
\begin{equation}
    \label{eq:gravitational time delay}
    t_{\mathrm{grav}}= - \int_0^{r_s} {\rm d}r\, \delta \phi_0(\bm{r}) D (r),
\end{equation}
where $D(r)$ is the linear growth factor, $\delta \phi_0$ is the potential fluctuation evaluated using the present-day matter field and we have used the weak field limit $\delta \phi \ll 1$. For a set of point masses $\phi = \sum_i G M_i/r_i$, but in a general density field $\delta \phi$ must be found by solving the Poisson equation
\begin{equation}
    \nabla^2\delta\phi(\bm{r}) = 4 \pi G \delta\rho(\bm{r}),
\end{equation}
where $\delta\rho$ is the total matter density fluctuation.
This is most readily solved in Fourier space. Thus determining time delays amounts to mapping out the three dimensional density field to at least the redshift of the source. We describe how we do this in \Cref{sec:density_inference}.

There exist several formulations of the Equivalence Principle, of varying strengths.
The WEP, our focus here, states that all freely-falling test objects follow the same trajectories given the same initial conditions, irrespective of their composition or structure.
This requires that all objects from a given source experience the \textit{same} time delay, regardless of their composition or energy.

Multiple astrophysical sources have been detected with precise timings of photons of different wavelengths. One contribution to the difference in the time of reception of these photons is a difference in gravitational time delay, which can be parametrised as a difference in the PPN parameter $\gamma$ between two wavelengths. Thus by comparing the measured time delays with the line-of-sight integral in \autoref{eq:gravitational time delay}, one can put constraints on how $\gamma$ changes with photon frequency, and thus on any violation of the WEP.

\subsection{Time delay angular power spectrum}
\label{sec:Time delay power spectrum}

The angular power spectrum of the gravitational time delay is \cite{Nusser_2016}
\begin{equation}
    C_{\ell} = \frac{2}{\pi}   \int {\rm d} k \, k^2 P_\phi \left( k \right) \left| \int_0^{r_{\rm s}} {\rm d} r \, D \left( r \right) j_{\ell} \left( k r \right) \right|^2,
\end{equation}
such that the mean-squared time delay is
\begin{equation}
    \label{eq:Cl to rms}
    \left< t_{\rm grav}^2 \right> = \sum_{\ell} \frac{2 \ell + 1}{4\pi} C_\ell,
\end{equation}
where $P_\phi$ is the power spectrum of the potential. For $k \ll 1 / r_{\rm s}$, the spherical Bessel function can be expanded as 
\begin{equation}
    j_{\ell} \left( x \right) = \frac{\sqrt \pi}{2^{\ell + 1} \Gamma \left( \ell + \frac{3}{2} \right)} x^{\ell} + \mathcal{O} \left( x ^{2 + \ell} \right),
\end{equation}
so that
\begin{equation}
    \begin{split}
	C_{\ell} =
	\frac{4}{\pi^3} & \frac{\sqrt \pi}{2^{\ell + 1} \Gamma \left( \ell + \frac{3}{2} \right)}
	\int \frac{{\rm d} k}{k} \left[ T_\phi \left( k \right )\right]^2A_{\rm s} k^{n_{\rm s} - 1} k^{2 \ell} \\
	& \times \left( \left| \int {\rm d} r \, D \left( r \right) r^{\ell} \right|^2 + \mathcal{O} \left( k ^{2 + 2\ell} \right) \right),
	\end{split}
\end{equation}
where $T_\phi$ is the potential transfer function and we assume a nearly scale-invariant primordial power spectrum
\begin{equation}
	\mathcal{P}_{\mathcal{R}}\left(k \right) = A_{\rm s} k^{n_{\rm s} - 1}.
\end{equation}
On super-Hubble scales, $T_\phi\left(k \right) \sim $ constant, so the smallest power of $k$ in the integral has an exponent $n_{\rm s} -2 + 2\ell$. For $\ell \geq 1$, this is $\geq n_{\rm s}$,  which, given that $n_{\rm s} \sim 0.97$ \cite{Planck_2018_I}, means that, for $\ell \geq 1$, the integral does not diverge at small $k$. However for $\ell=0$, the exponent is $n_{\rm s} - 2 < -1$, and so the integral diverges. We therefore see that the monopole diverges due to the contribution from large scales. This is not to say that there is an infinite time delay for a given universe, but that its variance across all possible universes is infinite. 

The problematic diverging monopole was first noted by \citet{Reischke_2021}, who showed how to circumvent this issue by computing the angular power spectrum and forecasted constraints of $\Delta \gamma < 10^{-15}$ using FRBs. Similarly, for our forward modelling approach in \Cref{sec:Methods}, we will only be able to predict angular fluctuations about the mean Shapiro time delay at a given redshift, and not its absolute value.

\subsection{Other contributions to the time delay}

We suppose that the time delay between frequencies $\nu_{i}$ and $\nu_{j}$ for a source at location $\bm{r}$ can be written as the sum of the following independent terms
\begin{equation}
    \label{eq:Time delay contributions}
	\Delta t_{i j} \left( \bm{r} \right) = \Delta \gamma_{i j} t_{\rm grav} 
	+ A \left( r \right) \left( \nu_{i}^{-2} - \nu_{j}^{-2} \right) + B_{i j}.
\end{equation}
Besides the Equivalence Principle-violating term, our time delay contains two other contributions. The first of these (containing $A\left(r\right)$) describes the dispersion due to electrons. The second ($B_{ij}$) represents the combination of an intrinsic time delay at the source and the instrument response, which we assume is independent of observed angle and redshift.

The mean of $A$ depends on the temporal evolution of the comoving electron density, $\bar{n}_{e,c}$, as 
\begin{equation}
	\bar{A} \left( r \right) = \frac{e^2}{2 \pi m_e} \frac{1}{4\pi\epsilon_0} \int_0^{z\left( r \right)} \frac{{\rm d} z^\prime}{H \left(z^\prime \right)} \left( 1 + z^\prime \right) \bar{n}_{e,c} \left( z^\prime \right),
\end{equation}
and fluctuations about this depend on fluctuations in the electron density and other relativistic effects \cite{Alonso_2021}. The comoving electron density can be modelled as
\begin{equation}
	\bar{n}_{e,c} \left( z \right) = \frac{3 H_0^2 \Omega_b}{8 \pi G m_p} \frac{x_e \left( z \right) \left( 1 + x_H \left( z \right) \right) }{2},
\end{equation}
where $x_e$ is the free electron fraction, $x_H$ is the hydrogen mass fraction, $\Omega_b$ is the baryon density fraction, and $m_p$ is the proton mass. $x_e$ is proportional to the fraction of electrons in the intergalactic medium, $f_{\rm IGM} \left( z \right)$, which slightly increases with redshift, from 0.8 for $z \lesssim 0.4$ to 0.9 at $z \gtrsim 1.5$ \cite{Meiksin_2009,Becker_2011,Shull_2012}. For simplicity, assuming $x_e = 1$ and $x_H = 0.75$, we find the contribution from the electron plasma to be
\begin{equation}
    \begin{split}
    A \left( r \right) & \left( \nu_{i}^{-2} - \nu_{j}^{-2} \right) 
    = \left[ \int_0^{z\left( r \right)} \frac{{\rm d} z^\prime}{E \left(z^\prime \right)} \left( 1 + z^\prime \right)\right]  \\
	& \times \left[ \left(\frac{E_i}{\rm keV} \right)^{-2} - \left(\frac{E_j}{\rm keV} \right)^{-2} \right] \times 5.6 \times 10^{-17} {\rm \, s},
	\end{split}
\end{equation}
where $E\left(z \right) \equiv H\left(z\right) / H_0$. Since in this work we will be considering Gamma Ray Bursts, we will find that this contribution is negligible for the probed frequencies, and thus we will neglect it in our analysis. If we were to consider radio bursts we would need to consider this term, as we would expect contributions of $\mathcal{O} \left(1 {\rm \, s} \right)$.

\section{Observational data}
\label{sec:Observational data}

The spectral lag data we use are a sample of the BATSE detections of GRBs catalogued by \citet{Hakkila_2007}. We make use of a set of sources compiled by \citet{Yu_2018}, where the four energy bins considered are sensitive to the ranges
Ch1: 25-60 keV, 
Ch2: 60-110 keV, 
Ch3: 110-325 keV and 
Ch4: $>$325 keV. 
These four bins result in up to 6 time delay pairs per source, $\Delta t_{ij}$, where $i,j$ label the channels used. Without loss of generality we define $i > j$. In the cases where no $\Delta t_{ij}$ is recorded for a source due to low signal-to-noise, we ignore that particular pair but still consider the others. These sources also have pseudo-redshifts calculated using the spectral peak energy-peak luminosity relation \cite{Yonetoku_2004}.

The physical mechanisms that result in the GRB spectral lag are unknown; the search for the nature of spectral lag is an ongoing research topic where most of the focus lies on investigating possible effects at the source \citep[see e.g.][]{Shen_2005,Lu_2006,Daigne_2003,Peng_2011,Du_2019,Lu_2018,Uhm_2016}. It has, for example, been shown that the effects of spectral lag can be recreated from simple source models utilising rapid bulk acceleration on relativistic jet shells \cite{Uhm_2016}. Therefore, when modelling the intrinsic contribution to these time delays in \Cref{sec:Noise model} we will have to rely on empirical models, as opposed to the ideal case where we can calibrate our noise model with simulations \citep{Bartlett_2021_HAGN}.

\section{Methods}
\label{sec:Methods}

In this section we use the large-scale structure information described in \Cref{sec:density_inference} to produce a source-by-source probabilistic forward model for the expected Shapiro time delay difference for a given $\Delta \gamma_{ij}$. Combining this with an empirical model describing other contributions to the measured time delays (noise) outlined in \Cref{sec:theory}, we calculate the likelihood function and constrain $\Delta \gamma_{ij}$ and the parameters describing the noise via a MCMC algorithm. The parameters which are fixed in this section are summarised in \autoref{tab:fixed_parameter_summary}.

\begin{table*}
    \caption{Parameters used to constrain the Equivalence Principle-violating contribution to the time delay. Above the horizontal line are the parameters used to forward model the time delay and below are the parameters passed to \textsc{multinest} in the MCMC analysis. In the final column we give the value chosen for each parameter, although we show in \Cref{sec:Discussion} that our results are unchanged for reasonable alternative values.}
    \label{tab:fixed_parameter_summary}
    \centering
    \begin{tabular}{l|l|l}
    \textbf{Parameter} & \textbf{Description} & \textbf{Value} \\
    \hline
    $L_{\rm box} \ / \ h^{-1} {\rm \, Mpc}$ & Side length of box used to reconstruct local density field. & 4000 \\
    $N_{\rm box}$ & Number of grid points per side of box used to reconstruct local density field. & 256 \\
    $k_0 \ / \ {\rm Mpc^{-1}}$ & Minimum wavenumber used to calculate long wavelength contribution. & $10^{-5}$ \\
    $\ell_{\rm max}$ & Maximum multipole used to compute time delay from angular power spectrum. & 2000 \\
    $N_{\rm side}$ & Resolution of \textsc{HEALPix} map used to calculate the monopole. & 64 \\
    $N_{\rm mon}$ & Number of redshifts between $z=0.1$ and $2$ used to calculate the monopole. & 20 \\
    $N_{\rm z, R}$ & Number of redshift points used to calculate the second term in \autoref{eq:split gravitational time delay}. & 512 \\
    $N_{\rm z, L}$ & Number of redshift points used to calculate long-wavelength time delay contributions. & 1024 \\
    $N_{\rm k, L}$ & Number of wavenumber points used to calculate unconstrained time delay contributions. & 512 \\
    $N_{\rm MC}$ & Number of Monte Carlo runs to get the distribution of time delays for the template signal & $10^3$ \\
    & for a given density field. & \\
    $N_{\rm bin}$ & Number of redshift bins to determine redshift evolution of noise model. & 5 \\
    $N_{\rm B}$ & Number of density field reconstructions sampled from the BORG chain. & 18 \\
    \hline
    \texttt{n\_live\_points} & Number of live points used in \textsc{multinest} sampling. & 800\\
    \texttt{importance\_nested\_sampling} & Whether to use importance nested sampling with \textsc{multinest}. & True\\
    \texttt{multimodal} & Whether to allow mode separation in \textsc{multinest}. & True\\
    \texttt{evidence\_tolerance} & Evidence tolerance for \textsc{multinest}. & 0.5\\
    \texttt{sampling\_efficiency} & Sampling efficiency for \textsc{multinest}. & 0.8\\
    \texttt{const\_efficiency\_mode} & Whether to use constant efficiency mode in \textsc{multinest}. & False \\
    \end{tabular}
\end{table*}

\subsection{Bayesian large scale structure inference}
\label{sec:density_inference}

This work builds upon previous results of applying the {\sc{BORG}} algorithm to the data of the SDSS-III/BOSS galaxy compilation \citep[see e.g.][]{BORG_1,BORG_2,BORG_3,BORG_4,Lavaux_2016}.

The {\sc{BORG}} algorithm is a fully probabilistic inference method aimed at reconstructing matter fields from galaxy observations. This algorithm incorporates a physical model for gravitational structure formation, enabling inference of the three-dimensional density field and the corresponding initial conditions at an earlier epoch from present observations.

Specifically the algorithm explores a large-scale structure posterior distribution consisting of a Gaussian prior for the initial density field at an initial cosmic scale factor of $a=10^{-3}$, linked to a Poissonian model of galaxy formation at a scale factor $a=1$ via second order Lagrangian perturbation theory \citep[for details see][]{BORG_2}. The model accurately describes one-, two- and three-point functions and represents very well higher-order statistics, as was calculated by e.g. \citep{MOUTARDE1991,BUCHERT1994,BOUCHET1995,SCOCCIMARRO2000,Leclercq_2013}. Thus {\sc{BORG}} naturally accounts for the filamentary structure of the cosmic web typically associated with higher-order statistics induced by nonlinear gravitational structure formation processes. The posterior distribution also accounts for the systematic and stochastic uncertainties encountered in cosmological surveys, including survey geometries, selection effects and noise.

Applying the {\sc{BORG}} algorithm to the SDSS-III/BOSS galaxy sample \cite{Eisenstein_2011,Lavaux_2019}, three-dimensional matter density fields have been inferred on a cubic Cartesian grid of a side length of $4000 \, h^{-1} {\rm Mpc}$ consisting of 256$^3$ equidistant voxels.
This results in a grid resolution of $\sim 15 \, h^{-1} {\rm Mpc}$. The inference assumes a standard $\Lambda$CDM model with the following set of cosmological parameters: $\Omega_m = 0.2889, \Omega_\Lambda = 0.7111, \Omega_b = 0.048597, h=0.6774, \sigma_8= 0.8159 , n_s = 0.9667$ \citep{Planck_2018_I}.

We also consider the 2M++ particle-mesh reconstruction of BORG \cite{Lavaux_2016, Jasche_Lavaux}, which again has 256$^3$ equidistant voxels, but a smaller side length of $677.7 {\rm \, Mpc}/h$ and thus a finer spatial resolution.\footnote{A set of 100 dark matter-only simulations based on the 2M++ reconstruction, dubbed \textsc{CSiBORG} \cite{Bartlett_2021_VS}, is also available. This could be used to access smaller scales because the initial conditions are augmented with white noise. We find in \Cref{sec:Calculating resolved} that these scales are unimportant, however, so we do not consider these simulations further.} In \Cref{sec:Calculating resolved} we compare how well we can determine the Shapiro time delay using the SDSS-III/BOSS vs 2M++ reconstructions.

\subsection{Calculating the Shapiro time delay}
\label{sec:Calculating resolved}

As discussed in \Cref{sec:Time delay power spectrum}, the variance of the mean Shapiro time delay at a given redshift across all possible universes diverges. We can rephrase our problem so that this is not an issue. Let us decompose the gravitational time delay into two parts
\begin{equation}
	t_{\rm grav} \left(\bm{r} \right) = t_0 \left( r \right) + \delta t_{\rm grav}  \left(\bm{r} \right),
\end{equation}
where $t_0 \left( r \right)$ 
is the mean Shapiro time delay across all angles at a given $r$ (i.e. the monopole), and $\delta t_{\rm grav}$ gives the fluctuation about this. These fluctuations can be decomposed into the sum of three terms
\begin{equation}
	\delta t_{\rm grav} \left( \bm{r} \right) = 
	\delta t_{\rm grav}^{\left( L \right)} \left( \bm{r} \right) + 
	\delta t_{\rm grav}^{\left( R \right)} \left( \bm{r} \right) + 
	\delta t_{\rm grav}^{\left( S \right)} \left( \bm{r} \right)
\end{equation}
corresponding to the long wavelength (L; $k < k_{\rm min}$), resolved (R; $k_{\rm min} \leq k <  k_{\rm max}$) and short wavelength (S; $k \geq k_{\rm max}$) contributions respectively. The `resolved' contribution can be determined using the inferred matter fields from \textsc{BORG}.

To determine the resolution and box size required to accurately reconstruct the time delay fluctuation, we compare the results of calculating the angular power spectrum with a finite resolution and box size to the continuous, infinite volume case. Obviously the latter is impossible in practice, but we approximate this limit by choosing a sufficiently small minimum ($10^{-5} {\rm \, Mpc^{-1}}$) and large maximum ($100 {\rm \, Mpc^{-1}}$) $k$. To approximate the finite volume result, we consider a box length $L_{\rm box}$ with $N_{\rm box}$ grid cells along each side, so we have minimum and maximum (non-zero) $k$ of
\begin{equation}
    \label{eq:kmin and kmax}
	k_{\rm min} = \frac{2\pi}{L_{\rm box}}, \quad k_{\rm max} = \frac{\pi \left( N_{\rm box} - 1 \right)}{L_{\rm box}}.
\end{equation}
Note that we do not include a factor of $\sqrt{3}$ in $k_{\rm max}$ as the sphere in $k$-space of radius $k_{\rm max}$ does not fully fit inside the first Brillouin zone if this is included. To mimic using this box, we top-hat filter the potential power spectrum, allowing modes between $k_{\rm min}$ and $k_{\rm max}$.  We calculate two quantities: (i) the $C_{\ell}$'s themselves, and (ii) the RMS fluctuation in the time delay using \autoref{eq:Cl to rms}, where we find $\ell_{\rm max} = 2000$ to be more than sufficient (halving this to $\ell_{\rm max} = 1000$ does not change the final result).

\begin{figure*}
	 \includegraphics[width=\textwidth]{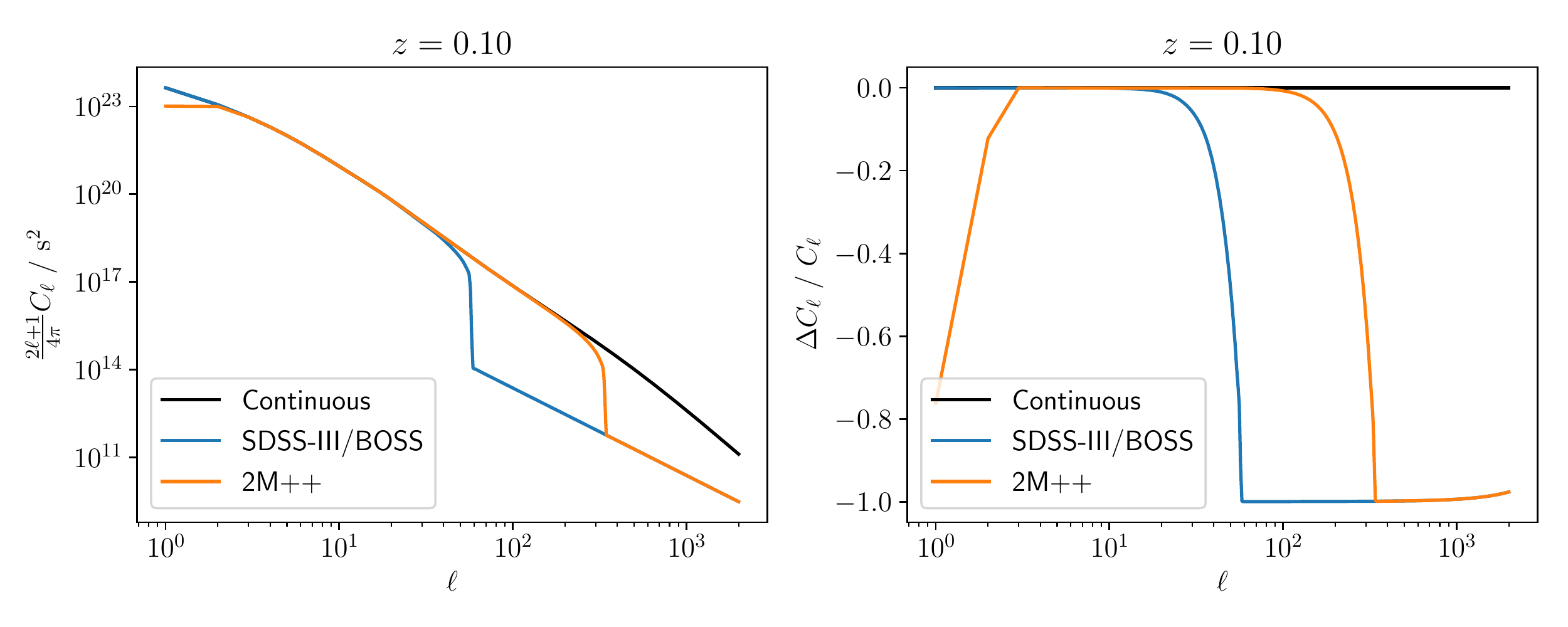}
	 \caption{\label{fig:Shapiro_Cls}The predicted Shapiro time delay fluctuation angular power spectrum for a source at $z=0.10$ calculated with the Core Cosmology Library. For the SDSS-III/BOSS and 2M++ calculations, we only include $k$ modes within the range sampled in the simulated volume due to the finite box length and resolution.  The right hand panel shows the fractional difference between these and the continuous case. Using SDSS-III/BOSS gives practically the same result for the time delay as the continuous case, whereas 2M++ underestimates this by $\sim 32$ per cent. The result is driven by the smallest $\ell$, making large box size far more important than high spatial resolution.}
\end{figure*}

In \autoref{fig:Shapiro_Cls} we compute the $C_{\ell}$'s for a source at redshift $z=0.10$ using the Core Cosmology Library \cite{CCL_2019}.
We compare the results from the continuous case with the SDSS-III/BOSS ($L_{\rm box} = 4000 h^{-1} {\rm \, Mpc}$, $N_{\rm box} = 256$) and 2M++ ($L_{\rm box} = 677.7 h^{-1} {\rm \, Mpc}$, $N_{\rm box} = 256$) reconstructions.
We note that the maximum distance a source can be from the observer in the latter two cases is $L_{\rm box} \sqrt{3} / 2$, which is $5.17 {\rm \, Gpc}$ for the SDSS-III/BOSS and $876 {\rm \, Mpc}$ for the 2M++ reconstruction, corresponding to redshifts of 1.9 and 0.21 respectively. Thus our test source at $z=0.10$ is within both boxes.

The continuous case gives the same result as SDSS-III/BOSS to $\sim 4$ parts in $10^5$ as the $C_{\ell}$'s only disagree at $\ell \gg 1$ while the result is dominated by low $\ell$. The lack of power at low $k$ results in the 2M++ box underestimating the low $\ell$ contribution, and thus the RMS time delay. The points of disagreement are as expected, since for this redshift, using the Limber approximation \cite{Lemos_2017} $\ell \sim k r$, we find that $k_{\rm min}$ and $k_{\rm max}$ correspond to multipoles
\begin{align}
	\ell_{\rm min}^{\rm SDSS-III/BOSS} &\sim 0.46, \quad \ell_{\rm max}^{\rm SDSS-III/BOSS} \sim 59 \\
	\ell_{\rm min}^{\rm 2M++} &\sim 2.7, \quad \ell_{\rm max}^{\rm 2M++} \sim 346,
\end{align}
which is approximately where we see the results diverging from the continuous case. Since the total time delay is dominated by low $\ell$, we conclude that a large box size is far more important than a fine resolution, justifying our choice of the SDSS-III/BOSS over 2M++ reconstruction. For simplicity, we will henceforth refer to the SDSS-III/BOSS \textsc{BORG} reconstruction simply as `\textsc{BORG}'.

Nonetheless, for sources at higher redshift, the density field not contained within the inferred volume contributes significantly to the gravitational time delay. For sources outside of the BORG volume, \autoref{eq:gravitational time delay} can be split into a contribution from the observer to the edge of the box, $r_{\rm b}$, and then from $r_{\rm b}$ to the source:
\begin{equation}
    \label{eq:split gravitational time delay}
    t_{\rm grav} = - \int_0^{r_{\rm b}} {\rm d}r\, \delta \phi_0(\bm{r}) D (r)
    - \int_{r_{\rm b}}^{r_s} {\rm d}r\, \delta \phi_0(\bm{r}) D (r).
\end{equation}
When we compute the first term (or \autoref{eq:gravitational time delay} for sources inside the box) we do not have a fluctuation, since the mean time delay at the upper limit of the integral is not zero. We therefore must find and subtract this mean. We do this by computing the time delays on \textsc{HEALPix}\footnote{\url{http://healpix.sf.net}} \citep{Zonca_2019,Gorski_2005} maps with $N_{\rm side} = 64$ at $N_{\rm mon} = 20$ logarithmically-spaced redshifts between $z=0.1$ and $z=2$. We then compute the monopoles of these maps with \textsc{healpy}. To determine the monopole at some intermediate redshift we linearly interpolate between the sampled redshifts.

\begin{figure}
	 \includegraphics[width=\columnwidth]{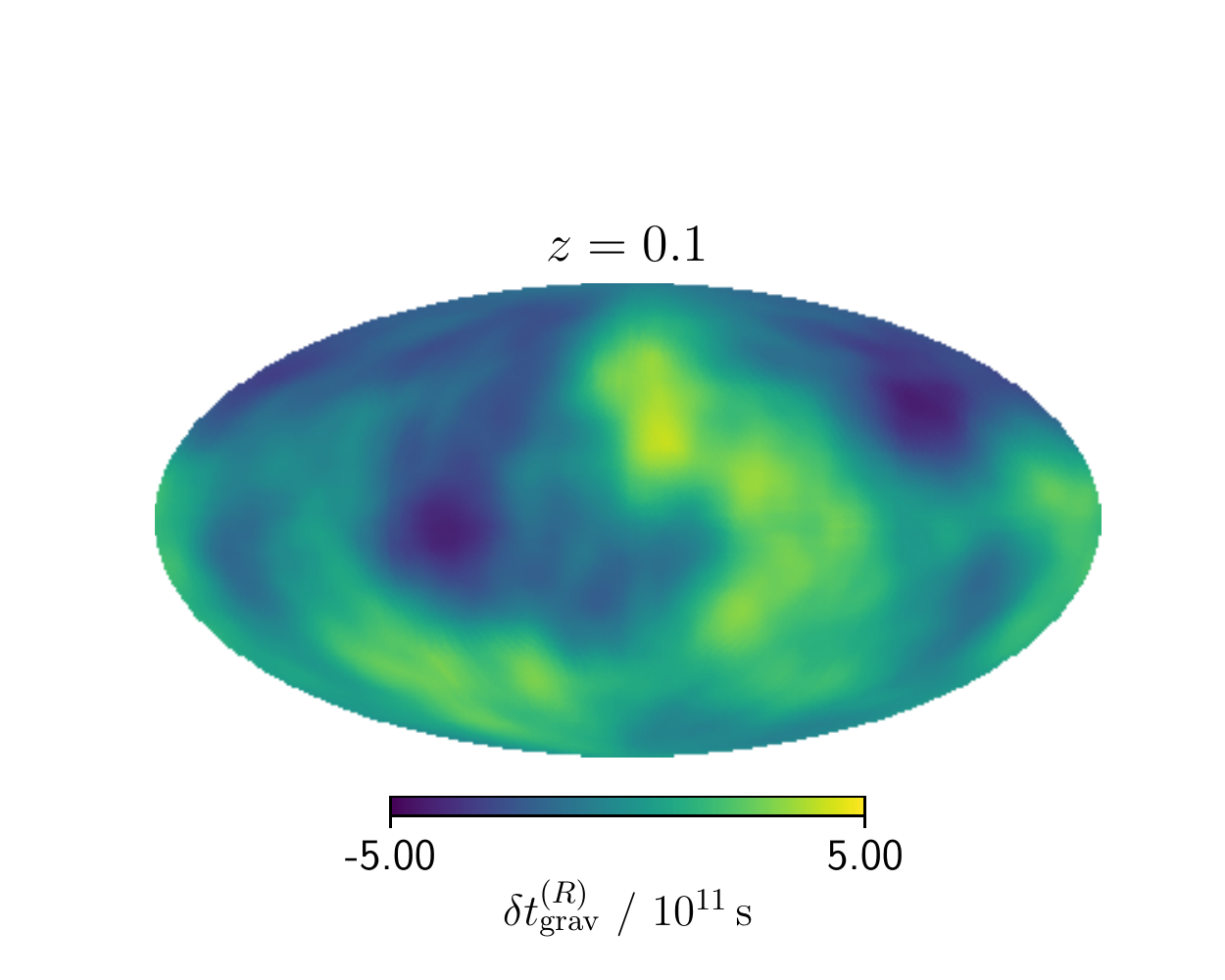}
	 \caption{\label{fig:fluctuation_plot}
	 Mollweide projection in equatorial coordinates of the ensemble mean of the time delay fluctuations at $z=0.1$ from resolved wavelengths. The typical scale is $\sim 10^{11} {\rm \, s}$ which, since the majority of observed time delays are $\mathcal{O}(0.1 {\rm \, s})$, indicates that constraints at least as tight as $|\Delta \gamma_{ij}| \lesssim 10^{-12}$ should be possible.
	 }
\end{figure}

The ensemble mean of the resulting time delay fluctuation map is plotted in \autoref{fig:fluctuation_plot}. One can see that the typical fluctuation from the resolved contribution is larger than $\sim 10^{11} {\rm \, s}$ and thus, given that the measured time delays are typically $\mathcal{O}(0.1 {\rm \,s})$, one would expect $| \Delta \gamma_{ij} |$ to be smaller than $\sim 10^{-12}$.

If the source is outside the volume, we now add an unconstrained contribution to the fluctuation (but not the monopole) drawn from a Gaussian of width given by \autoref{eq:Cl to rms}, but only using $k$ modes accessible to \textsc{BORG} (\autoref{eq:kmin and kmax}) and integrating between $r_{\rm b}$ and $r_{\rm s}$ using $N_{\rm z,R} = 512$ intermediate points. This corresponds to fluctuations due to the second term in \autoref{eq:split gravitational time delay}. We find $\ell_{\rm max} = 2000$ to be sufficient for this calculation.

To marginalise over the uncertainties in the pseudo-redshifts, angular position and unconstrained regions, we use simulation-based Bayesian forward modelling to create predictions from the statistical models. In particular, for a given density field from the converged part of the BORG MCMC chain, we take $N_{\rm MC} = 10^3$ Monte Carlo draws from the input distributions to build the likelihood.
For each iteration, we draw an angular location from Gaussian distributions characterised by the positional uncertainty given in \citep{Yu_2018} and a redshift from a two-sided Gaussian using the upper and lower uncertainties determined by \citep{Yonetoku_2004}. The time delay for a source at this position is then calculated for this density field. From \autoref{eq:Time delay contributions}, we see that the gravitational contribution to the time delay between any two frequencies is proportional to $\Delta \gamma_{ij}$. We therefore only need to run this Monte Carlo procedure once per source to construct a template signal with $\Delta \gamma_{ij} = 1$ since this can be trivially reintroduced as a scaling factor later.

\subsection{Monte Carlo modelling}

Now that we have $N_{\rm MC}$ samples per source and per BORG density field of our probabilistic model for $\delta t_{\rm grav}^{\left( R \right)}$, we must convert these samples into a distribution to use in our likelihood analysis. 

We model the samples as a Gaussian mixture model (GMM) \cite{scikit-learn}, where the likelihood function for some source $s$ and BORG density field $b$ is
\begin{equation}
    \begin{split}
	\mathcal{L}_{sb} \left( \delta t_{\rm grav}^{\left( R \right)} | \Delta \gamma_{ij} = 1 \right) &= \sum_\alpha \frac{w^{\left( \alpha \right)}_{sb}}{\sqrt{2 \pi}\tau^{\left( \alpha \right)}_{sb}} \\
	& \times  \exp \left[ - \frac{\left( \delta t_{\rm grav}^{\left( R \right)} - \lambda^{\left( \alpha \right)}_{sb}\right)^2}{2 {\tau^{\left( \alpha \right)}_{sb}}^2} \right],
	\end{split}
\end{equation}
where
\begin{equation}
	\sum_\alpha w_{\rm sb}^{\left( \alpha \right)} = 1, \quad w_{\rm sb}^{\left( \alpha \right)} \geq 0,
\end{equation}
and the sum runs over the Gaussians. The number of Gaussians is chosen to minimise the Bayesian Information Criterion (BIC)
\begin{equation}
\label{eq:BIC}
	{\rm BIC} = \mathcal{K} \log \mathcal{N} - 2 \log \hat{\mathcal{L}},
\end{equation}
for $\mathcal{K}$ model parameters, $\mathcal{N}=N_{\rm MC}$ data points, and maximum likelihood estimate $\hat{\mathcal{L}}$. Independent Gaussians are obtained for each source and BORG density field. To account for a different $\Delta \gamma_{ij}$, we must transform the means and widths of the Gaussians in the GMM as
\begin{equation}
    \tilde{\lambda}_{sbij}^{\left(\alpha \right)} = \Delta \gamma_{ij} \lambda_{sb}^{\left(\alpha\right)}, \quad
    \tilde{\tau}_{sbij}^{\left(\alpha \right)} = \left| \Delta \gamma_{ij} \right| \tau_{sb}^{\left(\alpha \right)}.
\end{equation}

\subsection{Adding larger scale modes}

Thus far we have only calculated the `resolved' contribution to the Shapiro time delay fluctuations. Since the large-$k$ contributions are negligible, we do not consider these further. To incorporate the long wavelength modes, we evaluate \autoref{eq:Cl to rms} for each source, where we integrate between $k_0 = 10^{-5} {\rm \, Mpc^{-1}}$ and $k_{\rm min}$ (\autoref{eq:kmin and kmax}) using $N_{\rm k, L} = 512$ intermediate points. We integrate up to the three-sigma redshift in the pseduo-redshift distribution with $N_{\rm z, L} = 1024$ points and again use $\ell_{\rm max} = 2000$. For source $s$ this gives the width of the distribution of long-wavelength contributions, $\xi_{s}$, which we assume to be Gaussian distributed since the density should also be Gaussian distributed on large scales. We must convolve this long-wavelength distribution with our GMM to get the total likelihood of $\delta t_{\rm grav}$,
\begin{equation}\label{eq:Like}
    \begin{split}
	\mathcal{L}_{sb} \left( \delta t_{\rm grav} \right) &= \sum_\alpha \frac{w^{\left( \alpha \right)}_{sb}}{\sqrt{2 \pi \left(\tilde{\tau}^{\left( \alpha \right)}_{sbij}{}^2 + \tilde{\xi}_{sij}^2 \right)}}  \\
	& \times \exp \left[ - \frac{\left( \delta t_{\rm grav} - \tilde{\lambda}^{\left( \alpha \right)}_{sb}\right)^2}{2 \left( \tilde{\tau}^{\left( \alpha \right)}_{sbij}{}^2 + \tilde{\xi}_{sij}^2 \right)} \right],
    \end{split}
\end{equation}
where
\begin{equation}
    \tilde{\xi}_{sij} \equiv \Delta \gamma_{ij} \xi_s.
\end{equation}

\subsection{Modelling the noise}
\label{sec:Noise model}

Given that we are neglecting the contribution from the electron plasma in \autoref{eq:Time delay contributions}, to determine the likelihood of an observed time delay, we must finally convolve Eq.~\ref{eq:Like} with the likelihoods for the intrinsic contribution and monopole terms. We assume that for each pair of frequencies the distribution of $B_{ij}$ can be written as the sum of $N_{\rm G}$ Gaussians,
\begin{equation}
    \mathcal{L} \left( B_{ij} \right) 
    = \sum_\beta \frac{\omega^{\left( \beta \right)}_{ij}}{\sqrt{2 \pi}\sigma^{\left( \beta \right)}_{ij}} \exp \left[ - \frac{\left( B_{ij} - \mu^{\left( \beta \right)}_{ij}\right)^2}{2 {\sigma^{\left( \beta \right)}_{ij}}^2} \right].
\end{equation}
where
\begin{equation}
	\sum_\beta \omega_{ij}^{\left( \beta \right)} = 1, \quad  \omega_{ij}^{\left( \beta \right)} \geq 0,
\end{equation}
and $\beta \in \{0, 1, \ldots, N_{\rm G} - 1 \}$. Without loss of generality, we define the Gaussians such that $\omega_{ij}^{\left( \beta \right)} \geq \omega_{ij}^{\left( \beta + 1 \right)}$.

For non-zero $\Delta \gamma_{ij}$, by binning the sources by redshift, one can constrain $t_0$ for each bin and a universal set of parameters describing $B_{ij}$, $\{ \omega^{\left( \beta \right)}_{ij}, \mu^{\left( \beta \right)}_{ij}, \sigma^{\left( \beta \right)}_{ij} \}$. However, we will find that $\Delta \gamma_{ij}$ is consistent with zero, which leaves $t_0$ completely unconstrained and thus a different approach is required. This is unfortunate since $t_0$ is the only parameter common to all different frequency pairs.

We therefore do not infer $\{ \mu^{\left( \beta \right)}_{ij} \}$, but rather 
\begin{equation}
    \tilde{\mu}^{\left( 0 \right)}_{ij} \left( z \right) \equiv \mu^{\left( 0 \right)}_{ij} + \Delta \gamma_{ij} t_0 \left( z \right),
\end{equation}
and $\{ \Delta \mu^{\left( \beta \right)}_{ij} \}$, where
\begin{equation}
    \Delta \mu^{\left( \beta \right)}_{ij} \equiv \mu^{\left( \beta \right)}_{ij} - \mu^{\left( 0 \right)}_{ij},
\end{equation}
such that
\begin{equation}
    \tilde{\mu}^{\left( \beta \right)}_{ij} \left( z \right) =
    \tilde{\mu}^{\left( 0 \right)}_{ij} \left( z \right) + \Delta \mu^{\left( \beta \right)}_{ij}.
\end{equation}

We bin our sources into $N_{\rm bin} = 5$ linearly spaced redshift bins between the minimum and maximum pseudo-redshifts in the sample and assume that all sources within each bin have the same $\tilde{\mu}^{\left( 0 \right)}_{ij} \left( z \right)$. Since we have now decoupled $\Delta \gamma_{ij}$ from the monopole term, our results are driven by the angular variation in the time delay and not the absolute value.

Finally, we must incorporate the errors on the measured $\Delta t_{ij}$. If these were independent of $\Delta t_{ij}$ then this could be incorporated into $\{ \sigma_{ij}^{\left( \beta \right)} \}$; however, longer lags tend to have larger errors \citep{Hakkila_2007}. We thus use the quoted measurement errors for each source and time delay pair, $\varepsilon_{sij}$, so that the likelihood for an observed time delay is 
\begin{equation}
    \begin{split}
    \mathcal{L}_{sb} \left( \Delta t_{ij} \right) &=
    \sum_{\alpha \beta}
    \frac{w^{\left( \alpha \right)}_{sb} \omega_{ij}^{\left( \beta \right)}}{\sqrt{2 \pi \left( {\tilde{\tau}^{\left( \alpha \right)}_{sbij}}{}^2  + \tilde{\xi}_{sij}^2 + \sigma_{ij}^{\left( \beta \right)}{}^2 + \varepsilon_{sij}^2\right) }} \\
    & \times \exp \left[ - \frac{\left( \Delta t_{ij} - \tilde{\lambda}^{\left( \alpha \right)}_{sbij} - \tilde{\mu}_{ij}^{\left( \beta \right)}\left( z_s \right)  \right)^2}{2 \left( {\tilde{\tau}^{\left( \alpha \right)}_{sbij}}{}^2 + \tilde{\xi}_{sij}^2 + \sigma_{ij}^{\left( \beta \right)}{}^2 + \varepsilon_{sij}^2 \right)} \right].
    \end{split}
\end{equation}

\subsection{Likelihood model}
\label{sec:Likelihood}

Treating each source as independent, the likelihood for our dataset $\mathcal{D}$ is then
\begin{equation}
    \mathcal{L}_{ij} \left(\mathcal{D}| \bm{\theta}\right) = 
    \prod_{s} \left(\frac{1}{N_{\rm B}} \sum_b \mathcal{L}_{sb} \left( \Delta t_{ij} \right) \right),
\end{equation}
where $\bm{\theta} \equiv \{ \Delta \gamma_{ij}, \tilde{\mu}_{ij}^{\left( 0\right)} \left( z \right), \Delta \mu_{ij}^{\left( \beta \right)}, \sigma_{ij}^{\left( \beta \right)}, \omega_{ij}^{\left( \beta \right)} \}$ and we use $N_{\rm B}=18$ BORG density field reconstructions, separated by 500 steps on the MCMC chain (approximately the autocorrelation length). Finally, given some prior on $\bm{\theta}$, $P(\bm{\theta})$, we use Bayes' theorem to obtain the posterior
\begin{equation}
	\mathcal{P}_{ij} \left(\bm{\theta}| \mathcal{D} \right) = \frac{\mathcal{L}_{ij} \left(\mathcal{D}|\bm{\theta} \right) P \left( \bm{\theta} \right) }{\mathcal{Z}_{ij} \left(\mathcal{D}\right)},
\end{equation}	
where the evidence, $\mathcal{Z}_{ij} (\mathcal{D})$, is the constant probability of the data for any $\bm{\theta}$.

We treat each pair of frequencies as independent and separately derive posteriors on $\Delta \gamma_{ij}$ and the noise parameters using \textsc{pymultinest} \cite{Multinest_1,Multinest_2,Multinest_3} with the settings given in \autoref{tab:fixed_parameter_summary}. The priors for all inferred parameters, given in \autoref{tab:infered_parameter_summary}, are found to be much broader than the posteriors for Gaussians with non-zero weights.

We run our inference with $N_{\rm G}$ in the range 1--5 (inclusive) to determine the posterior distributions for each noise model. To determine the appropriate $N_{\rm G}$, we compare the models by calculating the BIC (\autoref{eq:BIC}); the best-fitting model minimises this statistic. Since we now have access to the full posterior, as suggested by \citet{Handley_2019} we set $\mathcal{K} = \tilde{d}_{ij}$, where the Bayesian model dimensionality (BMD) is defined to be
\begin{equation}
    \label{eq:BMD}
    \frac{\tilde{d}_{ij}}{2} \equiv \int \mathcal{P}_{ij} \left(\bm{\theta}| \mathcal{D} \right) \left( \log \frac{\mathcal{P}_{ij} \left(\bm{\theta}| \mathcal{D}\right)}{P \left( \bm{\theta} \right)} \right)^2 {\rm d} \bm{\theta},
\end{equation}
and is computed using the \textsc{anesthetic} software package \citep{anesthetic}.
Although not as ``Bayesian'' as comparing the evidence, we prefer this statistic due to its insensitivity to the prior. We have deliberately set our priors very wide, making ratios of $\mathcal{Z}_{ij}$ difficult to interpret.

\begin{table}
    \caption{Model parameters describing the predicted signal and the empirical noise model. All priors are uniform in the range given.}
    \label{tab:infered_parameter_summary}
    \centering
    \begin{tabular}{c|c}
    Parameter & Prior \\
    \hline
    $ \Delta \gamma_{ij}$ & $\left[ -10^{-13}, 10^{-13} \right]$\\ 
    $\tilde{\mu}_{ij}^{\left( 0\right)} \left( z \right) / \, {\rm s}$ & $\left[ -10, 10 \right]$\\
    $\Delta \mu_{ij}^{\left( \beta \right)} / \, {\rm s}$ & $\left[ -10, 10 \right]$\\
    $\sigma_{ij}^{\left( \beta \right)} / \, {\rm s}$ & $\left[ 0, 10 \right]$\\
    $\omega_{ij}^{\left( \beta \right)}$ & $[0, 1], \quad \sum_\beta \omega_{ij}^{\left( \beta \right)} = 1, \quad \omega_{ij}^{\left( \beta \right)} \geq \omega_{ij}^{\left( \beta + 1 \right)}$\\
    \end{tabular}
\end{table}

\section{Results}
\label{sec:Results}

\begin{figure*}
     \centering
    \subfloat[]{
         \centering
         \includegraphics[width=0.45\textwidth]{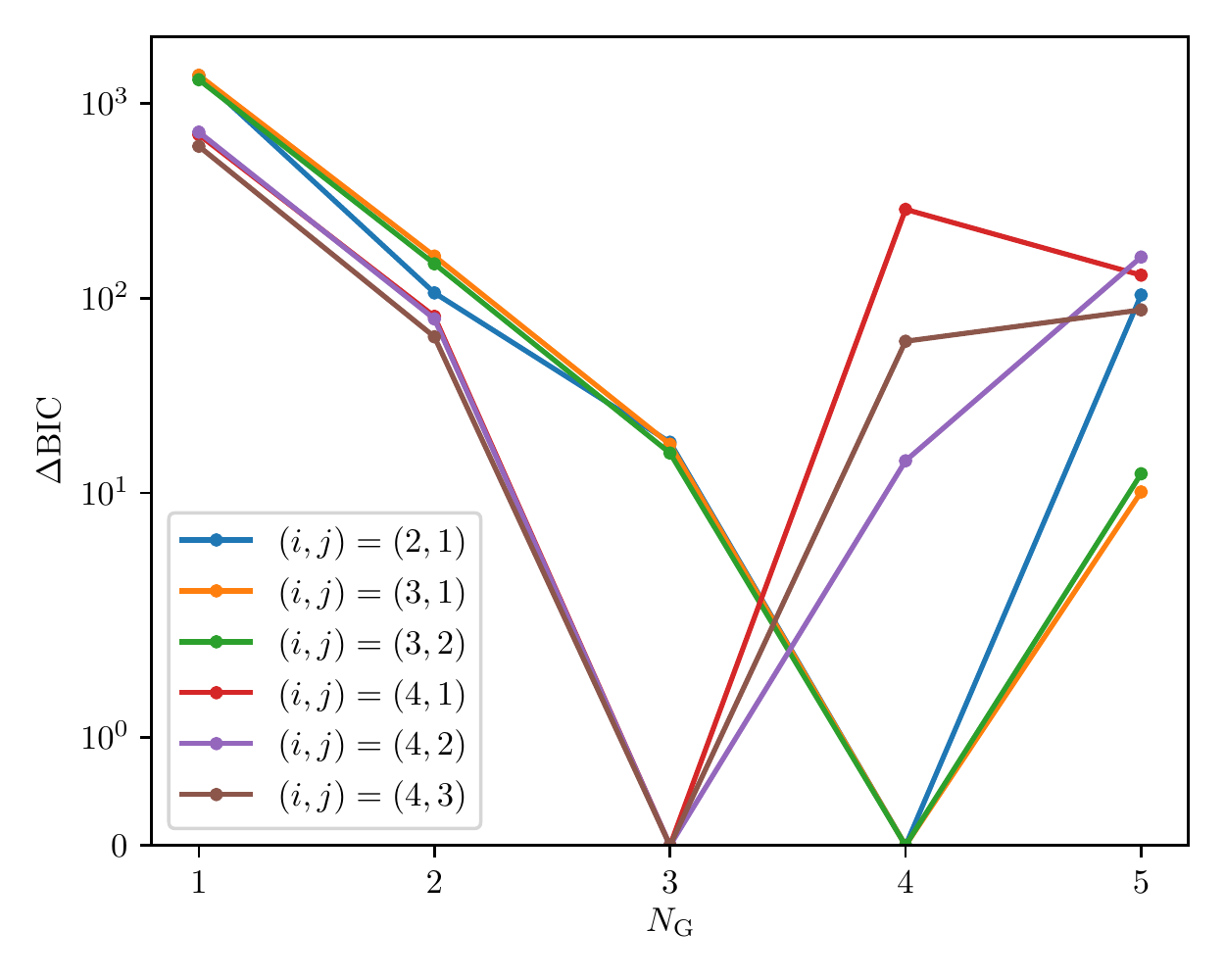}
         \label{subfig:BIC}
     }
     \subfloat[]{
         \centering
         \includegraphics[width=0.45\textwidth]{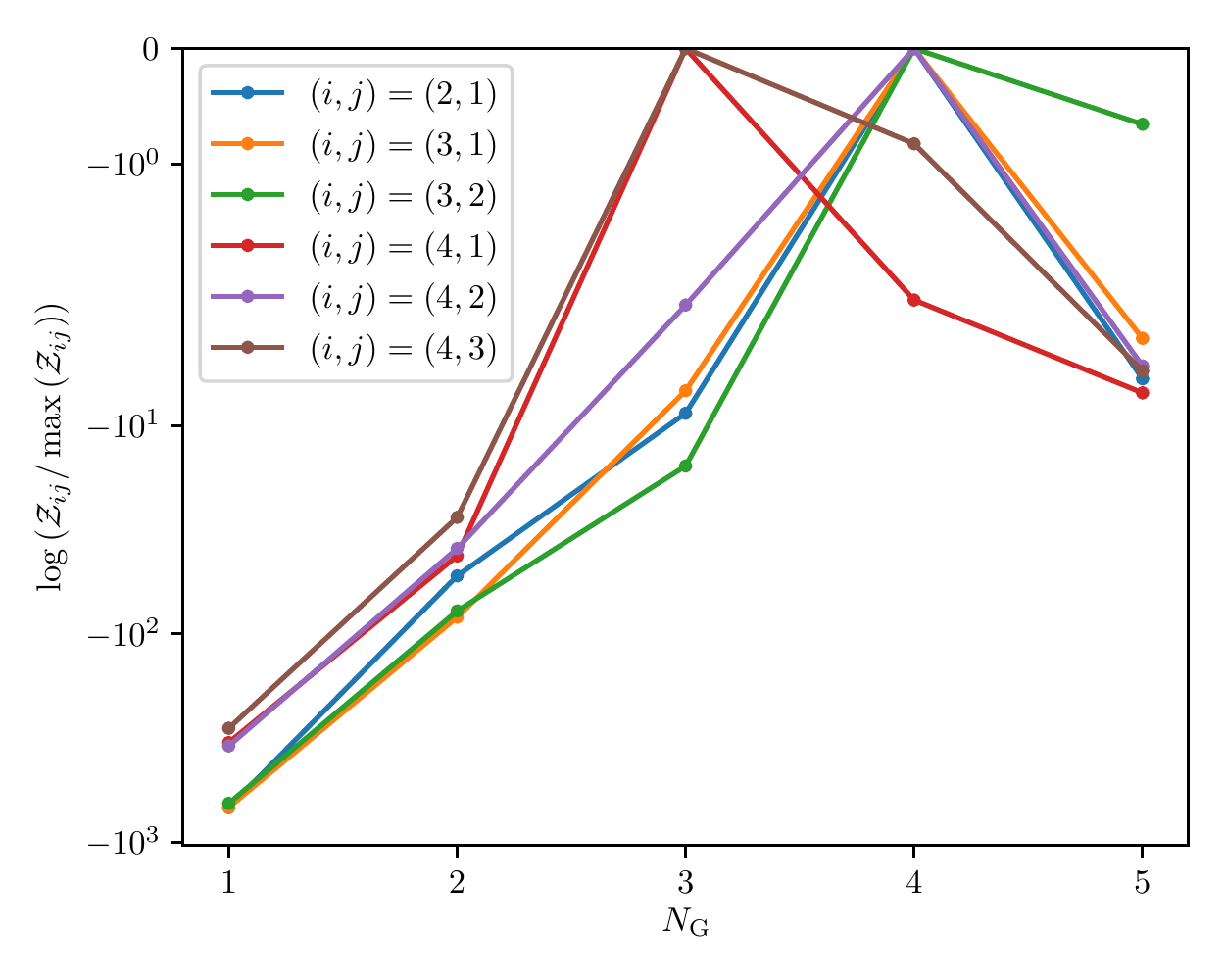}
         \label{subfig:Bayes}
     }
       \caption{Noise model comparison, including the Equivalence Principle-violating term, using \protect\subref{subfig:BIC} the BIC (\autoref{eq:BIC}) with the number of parameters of our model equal to the BMD (\autoref{eq:BMD}), and \protect\subref{subfig:Bayes} the Bayes factor. The best-fitting model should minimise the BIC and maximise the Bayes factor. For $i<4$ we see that $N_{\rm G}=4$ is the optimal noise model using either statistic. Although the Bayes factor prefers $N_{\rm G}=4$ for $(i,j)=(4,2)$, we adopt $N_{\rm G}=3$ for $i=4$ since this is favoured by the BIC, but find similar constraints on $\Delta \gamma_{42}$ for both $N_{\rm G}=3$ and $N_{\rm G}=4$.
         }
		\label{fig:Pick noise model}
\end{figure*}

In \autoref{fig:Pick noise model} we plot the BIC as a function of the number of Gaussian components in the noise model, $N_{\rm G}$, where we set the number of model parameters equal to the BMD. We find that the change in BIC is large when we use non-optimal $N_{\rm G}$, indicating that the best $N_{\rm G}$ is unambiguous. The same trend is found when we fix $\Delta \gamma_{ij}=0$ (i.e. fit the observations with just the noise model). For $N_{\rm G}=5$, we find that the weight of the lowest-weighted Gaussian, $\omega_{ij}^{\left( 4 \right)}$, is extremely small:
$\omega_{ij}^{\left( 4 \right)} < 0.03$ at 68\% confidence for all frequency pairs.
It is unsurprising then that the BIC increases for this case, since we are adding three new but negligible parameters to the noise model.
Similar results are found if we compare models using the Bayes factor; the only discrepancy is for $(i,j)=(4,2)$ where the BIC prefers $N_{\rm G}=3$ whereas the Bayes factor suggests that $N_{\rm G}=4$ is optimal. The constraints are similar for both models.

\begin{figure}
	 \includegraphics[width=\columnwidth]{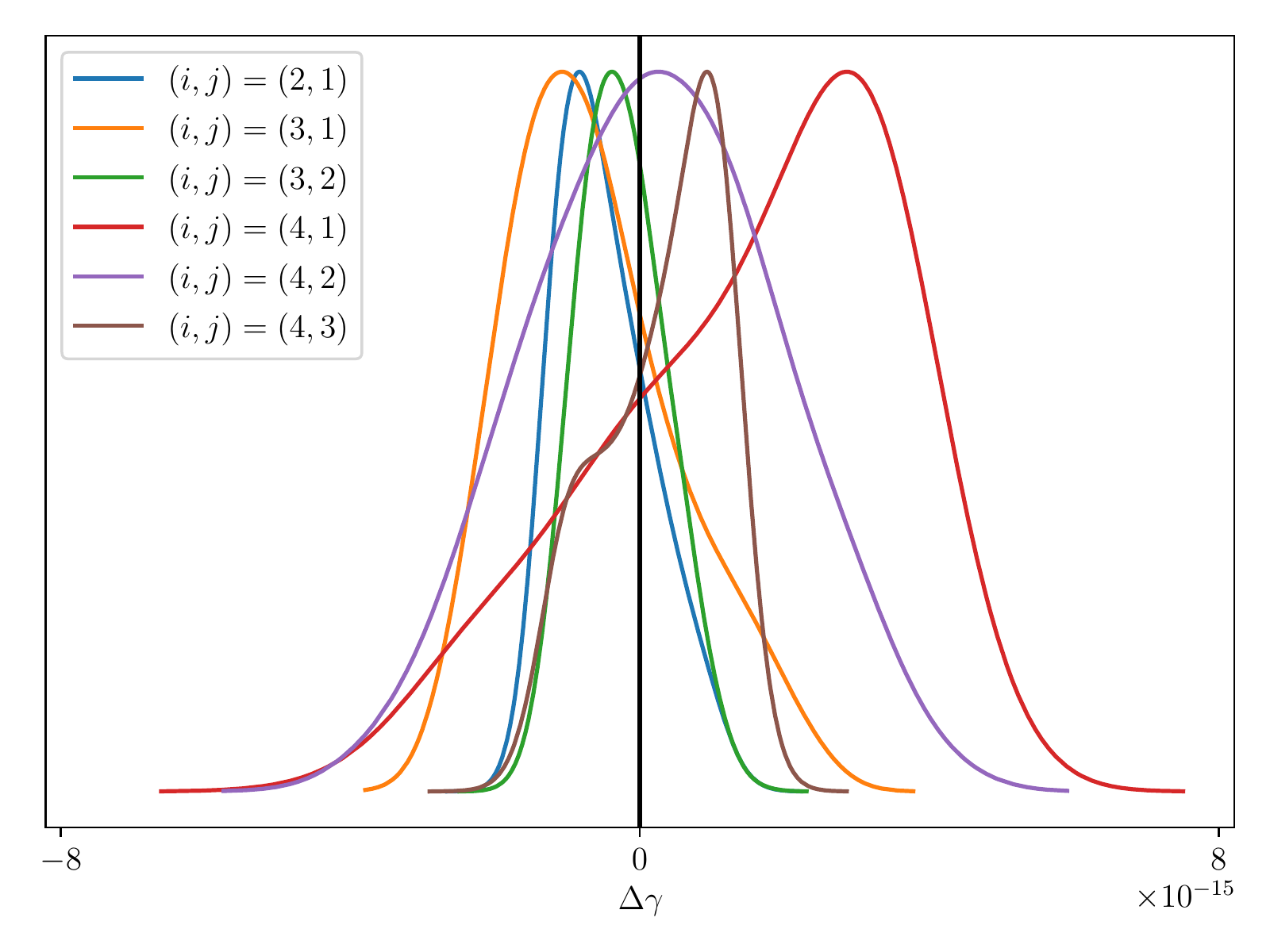}
	 \caption{\label{fig:Posterior 1D}Posteriors on $\Delta \gamma_{ij}$ for the different time delay pairs, marginalised over the noise parameters. These constraints use the noise models which minimise the BIC (\autoref{eq:BIC}).}
\end{figure}

\begin{figure*}
	 \includegraphics[width=\textwidth]{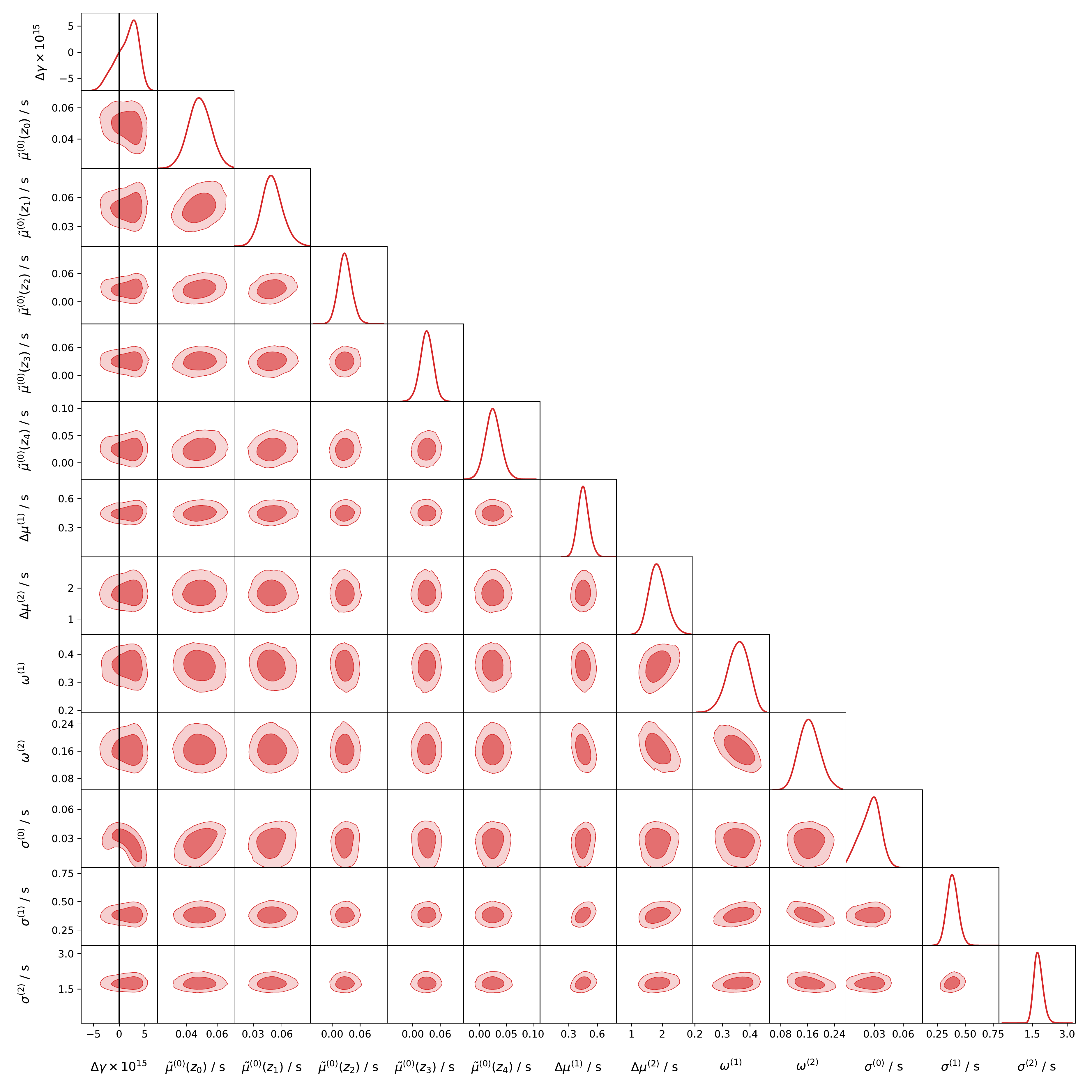}
	 \caption{\label{fig:Corner plot}Constraints on $\Delta \gamma_{ij}$ and the parameters describing other contributions to the time delay for $(i,j)=(4,1)$. The $ij$ indices have been suppressed for clarity since all parameters are only for this pair. The contours show the $1$ and $2\sigma$ confidence intervals.}
\end{figure*}

For the optimal $N_{\rm G}$ 
we plot the marginalised one-dimensional posteriors of $\Delta \gamma_{ij}$ in \autoref{fig:Posterior 1D}, and in \autoref{fig:Corner plot} we show the corner plot for the time delay pair with the weakest constraints on $\Delta \gamma_{ij}$. Since the unconstrained large scale contribution is modelled as a Gaussian of zero mean and finite width, if such scales alone were important then the one-dimensional posteriors would be symmetric about $\Delta \gamma_{ij} = 0$. That this is not true for all frequency pairs shows that our constraints on the density field are relevant.

For all time delay pairs we find $\Delta \gamma_{ij}$ to be consistent with zero. The $1\sigma$ constraints (defined to be half the difference between the 84\textsuperscript{th} and 16\textsuperscript{th} percentile) are tabulated in \autoref{tab:gamma_constraints}.
The tabulated results use the best-fit $N_{\rm G}$; however, we find that these change by no more than 25\% across the range $N_{\rm G}=3-5$.

We find that our weakest constraint is for the time delay between the highest and lowest energy channels ($(i,j)=(4,1)$), where we find $\Delta \gamma_{41} < \gcon$ at $1\sigma$ confidence. This is unsurprising because each pulse of a GRB's lightcurve is known to peak first at higher energy due to spectral evolution \citep{Kocevski_2003,Schaefer_2004,Ryde_2005,Hakkila_2011}, although the reasons for this are not fully understood \citep{Shen_2005,Lu_2006,Daigne_2003,Peng_2011,Du_2019,Lu_2018,Uhm_2016}. This results in the majority of observed time delays obeying $\Delta t_{ij} > 0 $ for $i > j$, so one expects the largest time delay for this pair and hence the weakest constraint.

\begin{table}
    \caption{The $1\sigma$ constraints on $\Delta \gamma_{ij}$ for the different time delay pairs.}
    \label{tab:gamma_constraints}
    \centering
    \begin{tabular}{c|c}
$(i,j)$ & Constraint on $\Delta \gamma_{ij} \times 10^{15}$ \\
\hline
$(2,1)$  &  0.68  \\
$(3,1)$  &  1.19  \\
$(3,2)$  &  0.62  \\
$(4,1)$  &  2.13  \\
$(4,2)$  &  1.72  \\
$(4,3)$  &  0.92  \\
    \end{tabular}
\end{table}

\begin{figure}
	 \includegraphics[width=\columnwidth]{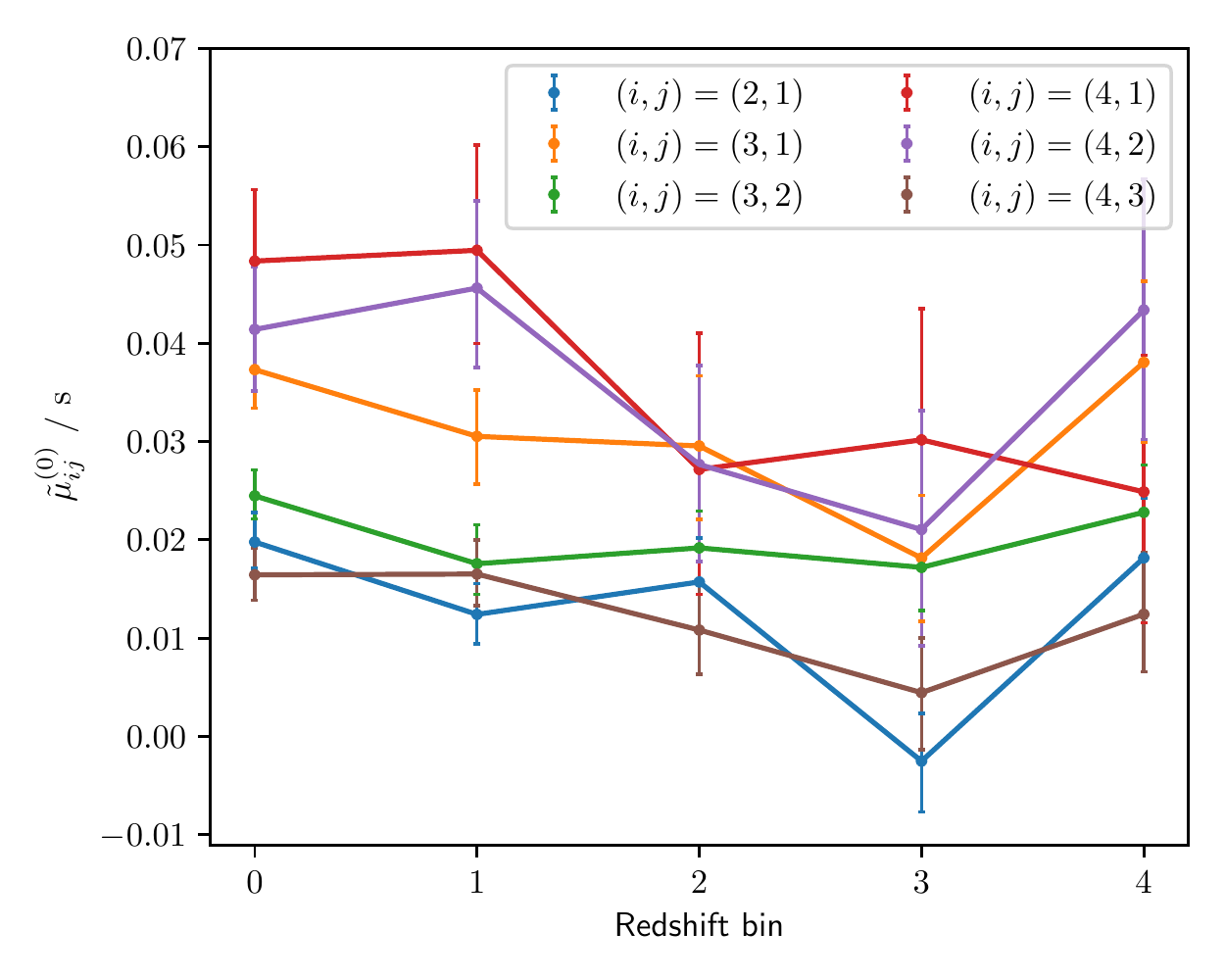}
	 \caption{\label{fig:redshift_mean_evolution}Evolution of the means of the highest-weighted Gaussian in the noise model, $\tilde{\mu}_{ij}^{\left( 0 \right)}$, with redshift, where it is assumed that all sources in the same redshift bin have the same $\tilde{\mu}_{ij}^{\left( 0 \right)}$. Any variation with redshift is comparable to the uncertainty on $\tilde{\mu}_{ij}^{\left( 0 \right)}$, as one would expect for a redshift-independent intrinsic time delay contribution, since $\Delta \gamma_{ij}$ is consistent with zero.}
\end{figure}

Given the form of our noise model, if $\Delta \gamma_{ij} = 0$ then one expects $\{ \tilde{\mu}_{ij}^{\left( 0 \right)} \}$ to be redshift independent. We check this in \autoref{fig:redshift_mean_evolution} and indeed find that for all pairs any variation in $\{ \tilde{\mu}_{ij}^{\left( 0 \right)} \}$ is comparable to its uncertainty, indicating that our assumption is reasonable. Furthermore, the time delay pair giving the weakest constraint ($(i,j)=(4,1)$) has noise parameters most suggestive of redshift evolution, as would be expected.

\section{Discussion}
\label{sec:Discussion}

\subsection{Systematic uncertainties}

Our probabilistic forward model is designed to propagate uncertainties in the source localisation and density field reconstruction, which we marginalise over via a MCMC algorithm. We use $N_{\rm MC} = 10^3$ Monte Carlo samples per source and per BORG density field to estimate the likelihood, but we also check that this is sufficiently large to fully sample the distributions we wish to marginalise over. Running the inference using $N_{\rm MC} = 500$ yields identical constraints on $\Delta \gamma_{ij}$ and the noise parameters as the fiducial case of $N_{\rm MC} = 10^3$, indicating that the number of samples is adequate.

\begin{figure}
	 \includegraphics[width=\columnwidth]{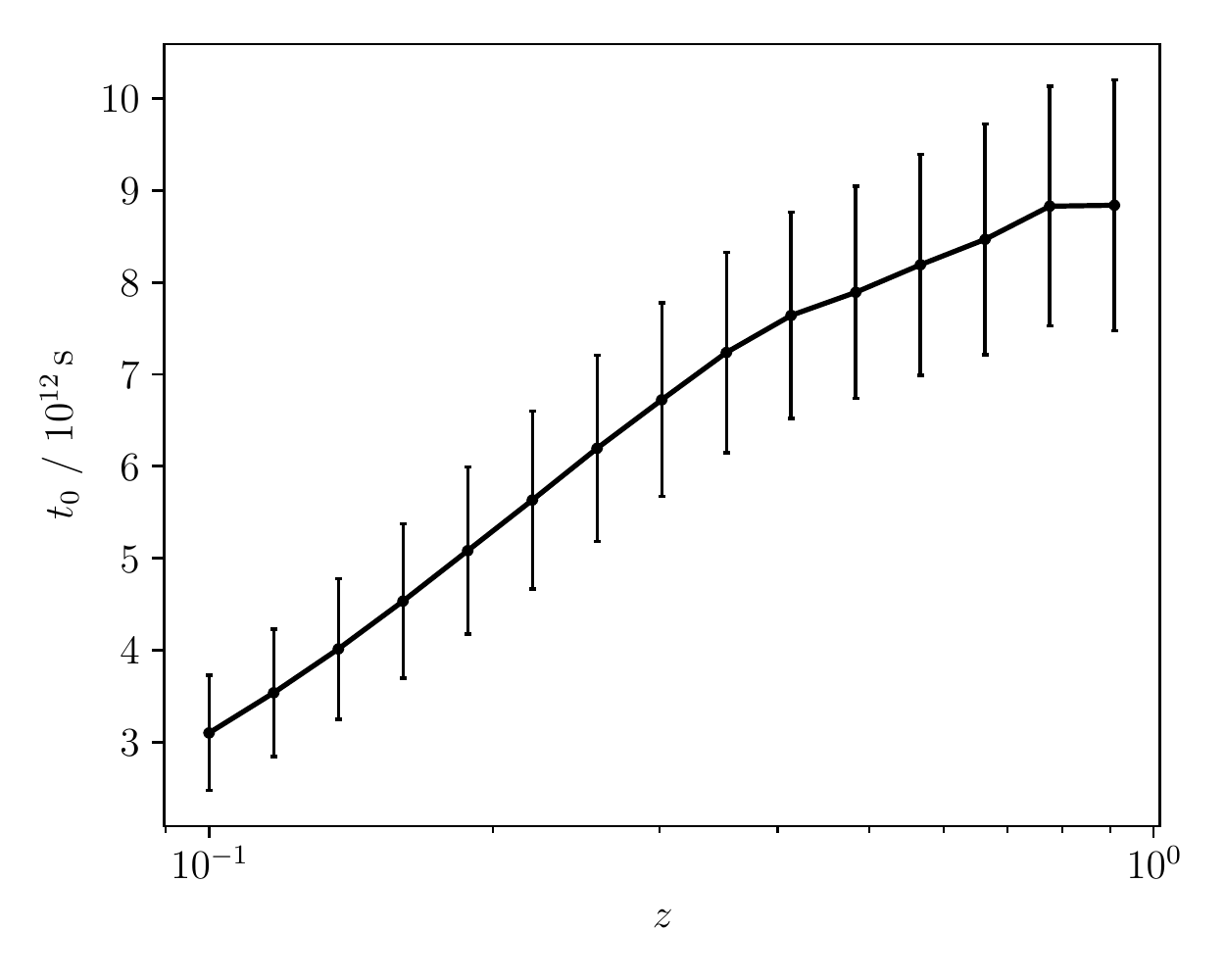}
	 \caption{\label{fig:monopole_parameters}Evolution of the ensemble mean of the monopole of Shapiro time delays, $t_0$, with redshift, $z$, calculated using wavelengths resolved within the BORG box. Note that the true monopole from all wavelengths cannot be predicted statistically (\Cref{sec:Time delay power spectrum}). The monopole is found to vary smoothly with redshift, suggesting that our interpolation procedure is reasonable. The ensemble mean is positive due to local massive structures.}
\end{figure}

In order to measure the fluctuations in the Shapiro time delay using BORG, we had to subtract the monopole at a given redshift. This involved sampling $N_{\rm mon}=20$ logarithmically spaced redshifts, where for each redshift we computed the monopole using a $N_{\rm side} = 64$ \textsc{HEALPix} map. 
In \autoref{fig:monopole_parameters} we plot the ensemble mean of the inferred monopole,
$t_0$, to subtract from the BORG contribution as a function of 
redshift, $z$, and see that $t_0$ is a smoothly varying function of $z$, indicating that $N_{\rm mon}=20$ should be sufficient. We find that if we were to use small values of $N_{\rm side}$ we would calculate the wrong monopole at high redshift, but that the values quickly converge with increasing $N_{\rm side}$. The maximum fractional difference between the calculated $t_0$ at $N_{\rm side}=32$ and $N_{\rm side}=64$ is $3 \times 10^{-3}$, indicating that our map's resolution is sufficient. Moreover, we run the inference again with $N_{\rm side}=32$ and $N_{\rm mon}=10$ and find $1\sigma$ constraints on $\Delta \gamma_{41}$ of $2.1\times10^{-15}$ and $2.0\times10^{-15}$ respectively, suggesting that our constraints are robust to these choices.

In \Cref{sec:Calculating resolved}, at each Monte Carlo iteration we drew a redshift from a two-sided Gaussian. One may be concerned that this is not the correct distribution.
To test the effect of the redshift uncertainty, we repeat the analysis but fix the redshift to the mean value; i.e., we assume zero redshift error. We find the constraints change by $\sim$10\% compared to our fiducial method, indicating that our constraints are not dominated by redshift uncertainty, and thus that the exact redshift distribution is not important.

Finally, to account for the monopole of the Shapiro time delay (which we are unable to predict or constrain by itself), we introduce a redshift dependence in our noise model, binning our sources into $N_{\rm bin}=5$ linearly spaced redshift bins. Repeating the inference with $N_{\rm bin}=1$ or $N_{\rm bin}=10$ changes our constraint to $\gamma_{41} < 1.8 \times 10^{-15}$ and $\gamma_{41} < 2.5 \times 10^{-15}$ respectively. This is as one would expect; the constraint tightens as we decrease the number of parameters in the noise model and weakens as this increases. However this change is only $0.1 {\rm \, dex}$, indicating that our constraint is relatively insensitive to the choice of $N_{\rm bin}$.

\subsection{Comparison with the literature}

The majority of previous attempts to constrain violations of the WEP via $\Delta \gamma_{ij}$ with GRBs \citep{Sivaram_1999,Gao_2015,Sang_2016,Wei_2016,Yu_2018}, FRBs \citep{Wei_2015,Luo_2016,Tingay_2016,Xing_2019,Wang_2020}, Supernovae \citep{Longo_1988,Krauss_1988}, Gravitational Waves \citep{Wu_2016,Kahya_2016,LIGO_2017,Liu_2017,Wei_2017,Shoemaker_2018,Boran_2018,Yao_2020,Yang_2020}, Blazar Flares \citep{Wang_2016,Wei_2016b,Boran_2019,Laha_2019,Wei_2019}, or Pulsars \citep{Yang_2016,Zhang_2017,Desai_2018,Leung_2018} have assumed that the gravitational potential is dominated by the contribution from the Milky Way and/or other massive objects such as the Laniakea supercluster. This has two major shortcomings for distant sources: first, the gravitational potential in \autoref{eq:gravitational time delay} should be a fluctuation about the cosmological mean (and thus can take either sign, unlike in the multiple-source approximation where it is strictly additive), and second, the long range behaviour of the gravitational potential means that we cannot neglect the large-scale distribution of mass \cite{Nusser_2016,Minazzoli_2019}. These studies claim that only including a few sources underestimates the Shapiro time delay, making the constraints obtained conservative \citep{Wei_2021}. However, since the potential can take either sign, this reasoning is incorrect.
Furthermore, the modelling of other contributions to the time delay in these studies are relatively simplistic compared to our analysis; for example, \citet{Yu_2018} assume all noise can be modelled as a single Gaussian, which we have shown to be a poor approximation. Not only have we considered more sophisticated noise models but also we have demonstrated that our constraints are robust to the choice of model.

As well as being more robust than previous work, our constraints are also stronger, and are comparable to the forecasts of \cite{Reischke_2021} for FRBs. The first attempts to include the cosmological contribution to the time delays yielded constraints weaker than $\Delta \gamma_{ij} < 10^{-13}$ \citep{Nusser_2016,Wang_2017}. Using the same sources as us and the point-mass methodology, \citet{Yu_2018} find $\Delta \gamma_{41} < 1.3 \times 10^{-13}$, which is a factor of $\sim$40 weaker than our constraints. As a consistency check, we re-run our inference for $(i,j)=(4,1)$ but (incorrectly) assume that the gravitational potential is dominated by the Laniakea supercluster (of mass $M = 10^{17} M_{\sun}$, at a distance $d=79 {\rm \, Mpc}$, with RA=10$^{\rm h}$32$^{\rm m}$ and Dec=-46$\degr$00$\arcmin$)
as is done by \citet{Yu_2018}. Also mimicking \citet{Yu_2018},
we now assume that the redshifts and angular positions of the sources have no uncertainty. We find constraints of $\Delta \gamma_{41} < 1.1 \times 10^{-14}$, which are tighter than \citep{Yu_2018}. This is due to our use of the optimal noise model with $N_{\rm G} = 3$; if we use $N_{\rm G} =1$ then our constraint is $\Delta \gamma_{41} < 1.1 \times 10^{-13}$, similar to that of \citep{Yu_2018}. One might have expected that increasing the number of Gaussian components describing the noise would weaken the constraint, not strengthen them. We find that for $N_{\rm G}=1$ the constraint is dominated by the few sources with the largest time delays.
This is because the WEP-violating term must account for the wide tails in the measured time delays, upweighting larger values of $|\Delta \gamma|$. When allowing $N_{\rm G}>1$, the broader Gaussians capture the tails instead, favouring smaller $|\Delta \gamma|$. This is preferable behaviour because the constraint on $\Delta \gamma$ should come from the angular correlation of the measured time delays with those predicted by BORG, rather than from the width of the measured time delay distribution itself.

The long Shapiro time delays for extragalactic sources result in tight constraints on $\Delta \gamma_{ij}$; however, one can only measure the time delay \textit{differences} and not the absolute time delays; we constrain $\Delta \gamma_{ij}$ and not $\gamma_i$. Therefore, although our constraints may appear tighter than Solar System measurements from the \textit{Cassini} spacecraft \citep{Bertotti_2003} or Very Long Baseline Interferometry \citep{Lambert_2009,Lambert_2011} of 
$\gamma - 1 = (2.1 \pm 2.3) \times 10^{-5}$
and 
$(-0.8 \pm 1.2) \times 10^{-4}$,
respectively, these have the advantage of constraining the PPN parameter itself and thus can differentiate between different theories that obey the Equivalence Principle. Nonetheless, our constraints indicate that the Equivalence Principle should be obeyed to within $\gcon$ for photons in the energy range $25-325 {\rm \, keV}$.

\subsection{Further applications}

In this work we have used the BORG reconstruction of the SDSS-III/BOSS galaxy compilation to predict the Shapiro time delay to GRBs. The mean pseudo-redshifts of the sources compiled by \citet{Yu_2018} is $\sim 2$, and thus we had to include unconstrained contributions to the calculation between the edge of the constrained volume and the sources (\autoref{eq:split gravitational time delay}). This, coupled with the importance of long wavelength modes (\Cref{sec:Calculating resolved}), suggests that our constraints could be improved by the next generation of galaxy surveys, such as with \textit{Euclid} \citep{Euclid_2011} or the Rubin Observatory \citep{LSST_2009}, which will be sensitive up to $z\sim2$. To estimate the potential improvement, we run the end-to-end inference assuming that we can accurately reconstruct the density field up to $z=1.26$ and to twice this redshift by generating Gaussian Random Fields. We find that the constraint improves by a factor of $\sim 30\%$ with this extra information. By combining the SDSS-III/BOSS and 2M++ reconstructions one could also reduce the uncertainty of the low-redshift part of the calculation, since 2M++ provides better constraints on the local density field.

A violation of the Equivalence Principle is not the only phenomenon which could lead to a delay between arrival times of photons of different energy. If the photon velocity is energy-dependent, then photons that travel more slowly will arrive later. If the photon has a non-zero rest mass, $m_\gamma$, then the speed of propagation increases with increasing energy
\begin{equation}
    v = \sqrt{1 - \frac{m_\gamma^2}{E^2}}
        \approx 1 - \frac{1}{2} \frac{m_\gamma^2}{E^2},
\end{equation}
($c\equiv1$), whereas in a quantum gravity scenario \citep{AmelinoCamelia_1998} one would expect a different scaling with energy
\begin{equation}
    v \sim 1 - \xi \frac{E}{E_{\rm QG}},
\end{equation}
where $\xi = \pm 1$ and $E_{\rm QG}$ is the quantum gravity energy scale, presumably around the Planck scale. In these cases, the Shapiro term is no longer dominant when computing the time delay, but the expected time delay for the massive photon is
\begin{equation}
    \label{eq:MP time delay}
    \Delta t_{ij} = \Delta v_{ij} \int_0^z \frac{{\rm d} z^\prime}{H \left( z^\prime \right) \left( 1 + z^\prime \right)^2},
\end{equation}
and for the quantum gravity case \citep{Jacob_2008}
\begin{equation}
    \label{eq:QG time delay}
    \Delta t_{ij} = \Delta v_{ij} \int_0^z \frac{\left( 1 + z^\prime \right)}{H \left( z^\prime \right)}{\rm d} z^\prime,
\end{equation}
where $\Delta v_{ij}$ is the difference in photon velocity between observed frequencies $\nu_i$ and $\nu_j$. Applying a similar methodology to that developed here would allow one to constrain on the photon mass and quantum gravity energy scale, and we will do so in future work.

We note that in this work we have used $\Delta\gamma_{ij}$ as a phenomenological parameter with which to quantify WEP violation of photons through \autoref{eq:gravitational time delay}. This is an arbitrary choice, and one could equally phrase equivalence principle violation as a difference in e.g. the gravitational constant \citep{Leonardo_2020}. Theories that predict a non-null signal in our test (e.g. massive photons) may have $\gamma=1$ in the usual PPN sense. In these cases, and neglecting cosmological redshift for illustrative purposes, one would equate $\Delta \gamma_{ij} = \Delta v_{ij}$ and thus for the above models our bound on $\Delta \gamma_{41}$ would correspond to $m_\gamma \lesssim 10^{-6} {\rm \, eV}$ and $E_{\rm QG} \gtrsim 10^{11} {\rm \, GeV}$, if $E_1 = 25 {\rm \, keV}$ and $E_4 = 325 {\rm \, keV}$. For such models the geometric term (\Cref{eq:MP time delay,eq:QG time delay}) dominates the time delay, so it is unsurpising that these bounds are weaker than the most stringent to date \citep{Xing_2019,Vasileiou_2013}.

Strong lensing of distant objects smears their light into an Einstein ring. The paths of photons observed across the ring trace out two cones that intersect at the lens plane. The relative time delays of these photons therefore contain information not only on the distribution of mass in the lens that sources a large part of the potential, but also on the path length itself which the time delay is proportional to. This length is a function of $H_0$, enabling this fundamental cosmological parameter to be constrained by measuring the time delays across the ring \cite{Holicow_2020, Birrer_2020}. As in this work, the modelling is potentially sensitive to mass further away from the geodesics that the photons follow than either the lens itself or the few additional massive objects modelled in \cite{Holicow_2020}. Provided one has sufficient spatial resolution to determine time delay differences across the Einstein ring, by using the time delay maps from constrained density fields one can test the validity of the external source assumptions.

Similarly, in GR photons and gravitational waves are predicted to follow the same geodesics. Recently, by considering the effect of four massive halos along the line of sight, it was shown \citep{Rubin_2020} at $> 5\sigma$ confidence that GW170817 \cite{GW170817} underwent gravitational lensing. As with the case of measuring $H_0$, one could use constrained density fields to determine the impact of mass away from the line of sight when computing the time delay.

In this work we have only considered the time delay \textit{differences} between different frequencies from a given source. The only cosmological regime in which the time delay itself may be directly measurable is the cosmic microwave background, where different delays between different regions of the sky imply varying times of recombination, and correspondingly varying temperature of the blackbody radiation \citep{Hu_2000}. Standard autocorrelation techniques cannot currently detect such a delay; however, cross-correlation with other fields appears to be a promising avenue \citep{Li_2019}. We leave it to future work to determine whether the large scale structure information from the \textsc{BORG} algorithm could afford a detection of this phenomenon.

\section{Conclusions}
\label{sec:Conclusions}

The PPN parameter $\gamma$ is the same for all particles at all energies if the WEP is obeyed. If there is a difference in $\gamma$, $\Delta \gamma_{ij}$, between two photon frequencies, $\nu_i$ and $\nu_j$, then there will be a spectral lag proportional to $\Delta \gamma_{ij}$ and the Shapiro time delay. Most previous attempts to calculate the Shapiro delay in order to constrain $\Delta \gamma_{ij}$ have assumed that the gravitational potential is dominated by only a few local sources and incorrectly argue that this produces conservative constraints.

In this work we constructed a source-by-source, Monte Carlo-based forward model for the Shapiro time delay from Gamma Ray Bursts detected by the BATSE satellite. We work in a cosmological context by combining the constrained local density field determined using the BORG algorithm with unconstrained, long-wavelength modes. Propagating uncertainties in the density field reconstruction via Monte Carlo sampling and marginalising over an empirical model characterising other contributions to the time delay, we derive constraints on $\Delta \gamma_{ij}$ between the four energy channels, and for all pairs find constraints at least as tight as $\Delta \gamma_{ij} < \gcon$ at $1\sigma$ confidence. These constraints are a factor $\sim$30 times tighter than previous results that use a cosmological model and $\sim$40 times tighter than if one neglects the cosmological contribution.

Our modelling is applicable to alternative multi-messenger probes of the WEP, although these may require different models for the other contributions to the time delays. For example, for FRBs one would need accurate maps of the electron density of the universe to forward model the contribution of scattering from the electron plasma, which depends on the integrated electron density (and is thus direction dependent) and scales as $\nu^{-2}$. Furthermore, cosmological calculations of the Shapiro time delay can determine the accuracy of lens modelling when considering time delays across the Einstein ring of a strongly lensed source or find use in the analysis of the cosmic microwave background. Identifying potential systematics in these probes of the Universe is vitally important, especially in the context of the Hubble tension \cite{Cosmology_Intertwined_1}.

\acknowledgements
{
We thank David Alonso and Jeremy Sakstein for useful discussions and Guilhem Lavaux for performing the SDSS-III/BOSS BORG inference.
D.J.B. is supported by STFC and Oriel College, Oxford. H.D. is supported by St John's College, Oxford. P.G.F. is supported by the STFC. H.D. and P.G.F. acknowledge financial support from ERC Grant No. 693024 and the Beecroft Trust. 
J.J. acknowledges support by the Swedish Research Council (VR) under the project 2020-05143 -- ``Deciphering the Dynamics of Cosmic Structure".
This work was done within the Aquila Consortium (\url{https://www.aquila-consortium.org/}).

Some of the results in this paper have been derived using the \textsc{healpy} and \textsc{HEALPix} packages.
}

\bibliographystyle{apsrev4-1}
\bibliography{references}

\begin{thebibliography}{96}%
\makeatletter
\providecommand \@ifxundefined [1]{%
 \@ifx{#1\undefined}
}%
\providecommand \@ifnum [1]{%
 \ifnum #1\expandafter \@firstoftwo
 \else \expandafter \@secondoftwo
 \fi
}%
\providecommand \@ifx [1]{%
 \ifx #1\expandafter \@firstoftwo
 \else \expandafter \@secondoftwo
 \fi
}%
\providecommand \natexlab [1]{#1}%
\providecommand \enquote  [1]{``#1''}%
\providecommand \bibnamefont  [1]{#1}%
\providecommand \bibfnamefont [1]{#1}%
\providecommand \citenamefont [1]{#1}%
\providecommand \href@noop [0]{\@secondoftwo}%
\providecommand \href [0]{\begingroup \@sanitize@url \@href}%
\providecommand \@href[1]{\@@startlink{#1}\@@href}%
\providecommand \@@href[1]{\endgroup#1\@@endlink}%
\providecommand \@sanitize@url [0]{\catcode `\\12\catcode `\$12\catcode
  `\&12\catcode `\#12\catcode `\^12\catcode `\_12\catcode `\%12\relax}%
\providecommand \@@startlink[1]{}%
\providecommand \@@endlink[0]{}%
\providecommand \url  [0]{\begingroup\@sanitize@url \@url }%
\providecommand \@url [1]{\endgroup\@href {#1}{\urlprefix }}%
\providecommand \urlprefix  [0]{URL }%
\providecommand \Eprint [0]{\href }%
\providecommand \doibase [0]{http://dx.doi.org/}%
\providecommand \selectlanguage [0]{\@gobble}%
\providecommand \bibinfo  [0]{\@secondoftwo}%
\providecommand \bibfield  [0]{\@secondoftwo}%
\providecommand \translation [1]{[#1]}%
\providecommand \BibitemOpen [0]{}%
\providecommand \bibitemStop [0]{}%
\providecommand \bibitemNoStop [0]{.\EOS\space}%
\providecommand \EOS [0]{\spacefactor3000\relax}%
\providecommand \BibitemShut  [1]{\csname bibitem#1\endcsname}%
\let\auto@bib@innerbib\@empty
\bibitem [{\citenamefont {{Shapiro}}(1964)}]{Shapiro_1964}%
  \BibitemOpen
  \bibfield  {author} {\bibinfo {author} {\bibfnamefont {I.~I.}\ \bibnamefont
  {{Shapiro}}},\ }\href {\doibase 10.1103/PhysRevLett.13.789} {\bibfield
  {journal} {\bibinfo  {journal} {\prl}\ }\textbf {\bibinfo {volume} {13}},\
  \bibinfo {pages} {789} (\bibinfo {year} {1964})}\BibitemShut {NoStop}%
\bibitem [{\citenamefont {{Bertotti}}\ \emph {et~al.}(2003)\citenamefont
  {{Bertotti}}, \citenamefont {{Iess}},\ and\ \citenamefont
  {{Tortora}}}]{Bertotti_2003}%
  \BibitemOpen
  \bibfield  {author} {\bibinfo {author} {\bibfnamefont {B.}~\bibnamefont
  {{Bertotti}}}, \bibinfo {author} {\bibfnamefont {L.}~\bibnamefont {{Iess}}},
  \ and\ \bibinfo {author} {\bibfnamefont {P.}~\bibnamefont {{Tortora}}},\
  }\href {\doibase 10.1038/nature01997} {\bibfield  {journal} {\bibinfo
  {journal} {\nat}\ }\textbf {\bibinfo {volume} {425}},\ \bibinfo {pages} {374}
  (\bibinfo {year} {2003})}\BibitemShut {NoStop}%
\bibitem [{\citenamefont {{Will}}\ and\ \citenamefont
  {{Nordtvedt}}(1972)}]{Will_1972}%
  \BibitemOpen
  \bibfield  {author} {\bibinfo {author} {\bibfnamefont {C.~M.}\ \bibnamefont
  {{Will}}}\ and\ \bibinfo {author} {\bibfnamefont {J.}~\bibnamefont
  {{Nordtvedt}}, \bibfnamefont {Kenneth}},\ }\href {\doibase 10.1086/151754}
  {\bibfield  {journal} {\bibinfo  {journal} {\apj}\ }\textbf {\bibinfo
  {volume} {177}},\ \bibinfo {pages} {757} (\bibinfo {year}
  {1972})}\BibitemShut {NoStop}%
\bibitem [{\citenamefont {{Wong}}\ \emph {et~al.}(2020)\citenamefont {{Wong}}
  \emph {et~al.}}]{Holicow_2020}%
  \BibitemOpen
  \bibfield  {author} {\bibinfo {author} {\bibfnamefont {K.~C.}\ \bibnamefont
  {{Wong}}} \emph {et~al.},\ }\href {\doibase 10.1093/mnras/stz3094} {\bibfield
   {journal} {\bibinfo  {journal} {\mnras}\ }\textbf {\bibinfo {volume}
  {498}},\ \bibinfo {pages} {1420} (\bibinfo {year} {2020})}\BibitemShut
  {NoStop}%
\bibitem [{\citenamefont {{Birrer}}\ \emph {et~al.}(2020)\citenamefont
  {{Birrer}} \emph {et~al.}}]{Birrer_2020}%
  \BibitemOpen
  \bibfield  {author} {\bibinfo {author} {\bibfnamefont {S.}~\bibnamefont
  {{Birrer}}} \emph {et~al.},\ }\href {\doibase 10.1051/0004-6361/202038861}
  {\bibfield  {journal} {\bibinfo  {journal} {\aap}\ }\textbf {\bibinfo
  {volume} {643}},\ \bibinfo {eid} {A165} (\bibinfo {year} {2020})}\BibitemShut
  {NoStop}%
\bibitem [{\citenamefont {{Luo}}\ \emph {et~al.}(2016)\citenamefont {{Luo}},
  \citenamefont {{Zhang}}, \citenamefont {{Wei}},\ and\ \citenamefont
  {{Wu}}}]{Luo_2016}%
  \BibitemOpen
  \bibfield  {author} {\bibinfo {author} {\bibfnamefont {Z.-X.}\ \bibnamefont
  {{Luo}}}, \bibinfo {author} {\bibfnamefont {B.}~\bibnamefont {{Zhang}}},
  \bibinfo {author} {\bibfnamefont {J.-J.}\ \bibnamefont {{Wei}}}, \ and\
  \bibinfo {author} {\bibfnamefont {X.-F.}\ \bibnamefont {{Wu}}},\ }\href
  {\doibase 10.1016/j.jheap.2016.04.001} {\bibfield  {journal} {\bibinfo
  {journal} {Journal of High Energy Astrophysics}\ }\textbf {\bibinfo {volume}
  {9}},\ \bibinfo {pages} {35} (\bibinfo {year} {2016})}\BibitemShut {NoStop}%
\bibitem [{\citenamefont {{LIGO Scientific Collaboration}}\ and\ \citenamefont
  {{Virgo Collaboration}}(2017)}]{GW170817}%
  \BibitemOpen
  \bibfield  {author} {\bibinfo {author} {\bibnamefont {{LIGO Scientific
  Collaboration}}}\ and\ \bibinfo {author} {\bibnamefont {{Virgo
  Collaboration}}},\ }\href {\doibase 10.1103/PhysRevLett.119.161101}
  {\bibfield  {journal} {\bibinfo  {journal} {\prl}\ }\textbf {\bibinfo
  {volume} {119}},\ \bibinfo {eid} {161101} (\bibinfo {year}
  {2017})}\BibitemShut {NoStop}%
\bibitem [{\citenamefont {{Nusser}}(2016)}]{Nusser_2016}%
  \BibitemOpen
  \bibfield  {author} {\bibinfo {author} {\bibfnamefont {A.}~\bibnamefont
  {{Nusser}}},\ }\href {\doibase 10.3847/2041-8205/821/1/L2} {\bibfield
  {journal} {\bibinfo  {journal} {\apjl}\ }\textbf {\bibinfo {volume} {821}},\
  \bibinfo {eid} {L2} (\bibinfo {year} {2016})}\BibitemShut {NoStop}%
\bibitem [{\citenamefont {{Minazzoli}}\ \emph {et~al.}(2019)\citenamefont
  {{Minazzoli}}, \citenamefont {{Johnson-McDaniel}},\ and\ \citenamefont
  {{Sakellariadou}}}]{Minazzoli_2019}%
  \BibitemOpen
  \bibfield  {author} {\bibinfo {author} {\bibfnamefont {O.}~\bibnamefont
  {{Minazzoli}}}, \bibinfo {author} {\bibfnamefont {N.~K.}\ \bibnamefont
  {{Johnson-McDaniel}}}, \ and\ \bibinfo {author} {\bibfnamefont
  {M.}~\bibnamefont {{Sakellariadou}}},\ }\href {\doibase
  10.1103/PhysRevD.100.104047} {\bibfield  {journal} {\bibinfo  {journal}
  {\prd}\ }\textbf {\bibinfo {volume} {100}},\ \bibinfo {eid} {104047}
  (\bibinfo {year} {2019})}\BibitemShut {NoStop}%
\bibitem [{\citenamefont {{Jasche}}\ and\ \citenamefont
  {{Wandelt}}(2012)}]{BORG_1}%
  \BibitemOpen
  \bibfield  {author} {\bibinfo {author} {\bibfnamefont {J.}~\bibnamefont
  {{Jasche}}}\ and\ \bibinfo {author} {\bibfnamefont {B.~D.}\ \bibnamefont
  {{Wandelt}}},\ }\href {\doibase 10.1111/j.1365-2966.2012.21423.x} {\bibfield
  {journal} {\bibinfo  {journal} {\mnras}\ }\textbf {\bibinfo {volume} {425}},\
  \bibinfo {pages} {1042} (\bibinfo {year} {2012})}\BibitemShut {NoStop}%
\bibitem [{\citenamefont {{Jasche}}\ and\ \citenamefont
  {{Wandelt}}(2013)}]{BORG_2}%
  \BibitemOpen
  \bibfield  {author} {\bibinfo {author} {\bibfnamefont {J.}~\bibnamefont
  {{Jasche}}}\ and\ \bibinfo {author} {\bibfnamefont {B.~D.}\ \bibnamefont
  {{Wandelt}}},\ }\href {\doibase 10.1093/mnras/stt449} {\bibfield  {journal}
  {\bibinfo  {journal} {\mnras}\ }\textbf {\bibinfo {volume} {432}},\ \bibinfo
  {pages} {894} (\bibinfo {year} {2013})}\BibitemShut {NoStop}%
\bibitem [{\citenamefont {{Jasche}}\ \emph {et~al.}(2010)\citenamefont
  {{Jasche}}, \citenamefont {{Kitaura}}, \citenamefont {{Wandelt}},\ and\
  \citenamefont {{En{\ss}lin}}}]{BORG_3}%
  \BibitemOpen
  \bibfield  {author} {\bibinfo {author} {\bibfnamefont {J.}~\bibnamefont
  {{Jasche}}}, \bibinfo {author} {\bibfnamefont {F.~S.}\ \bibnamefont
  {{Kitaura}}}, \bibinfo {author} {\bibfnamefont {B.~D.}\ \bibnamefont
  {{Wandelt}}}, \ and\ \bibinfo {author} {\bibfnamefont {T.~A.}\ \bibnamefont
  {{En{\ss}lin}}},\ }\href {\doibase 10.1111/j.1365-2966.2010.16610.x}
  {\bibfield  {journal} {\bibinfo  {journal} {\mnras}\ }\textbf {\bibinfo
  {volume} {406}},\ \bibinfo {pages} {60} (\bibinfo {year} {2010})}\BibitemShut
  {NoStop}%
\bibitem [{\citenamefont {{Jasche}}\ \emph {et~al.}(2015)\citenamefont
  {{Jasche}}, \citenamefont {{Leclercq}},\ and\ \citenamefont
  {{Wandelt}}}]{BORG_4}%
  \BibitemOpen
  \bibfield  {author} {\bibinfo {author} {\bibfnamefont {J.}~\bibnamefont
  {{Jasche}}}, \bibinfo {author} {\bibfnamefont {F.}~\bibnamefont
  {{Leclercq}}}, \ and\ \bibinfo {author} {\bibfnamefont {B.~D.}\ \bibnamefont
  {{Wandelt}}},\ }\href {\doibase 10.1088/1475-7516/2015/01/036} {\bibfield
  {journal} {\bibinfo  {journal} {\jcap}\ }\textbf {\bibinfo {volume} {2015}},\
  \bibinfo {eid} {036} (\bibinfo {year} {2015})}\BibitemShut {NoStop}%
\bibitem [{\citenamefont {{Lavaux}}\ and\ \citenamefont
  {{Jasche}}(2016)}]{Lavaux_2016}%
  \BibitemOpen
  \bibfield  {author} {\bibinfo {author} {\bibfnamefont {G.}~\bibnamefont
  {{Lavaux}}}\ and\ \bibinfo {author} {\bibfnamefont {J.}~\bibnamefont
  {{Jasche}}},\ }\href {\doibase 10.1093/mnras/stv2499} {\bibfield  {journal}
  {\bibinfo  {journal} {\mnras}\ }\textbf {\bibinfo {volume} {455}},\ \bibinfo
  {pages} {3169} (\bibinfo {year} {2016})}\BibitemShut {NoStop}%
\bibitem [{\citenamefont {{Planck Collaboration}}(2020)}]{Planck_2018_I}%
  \BibitemOpen
  \bibfield  {author} {\bibinfo {author} {\bibnamefont {{Planck
  Collaboration}}},\ }\href {\doibase 10.1051/0004-6361/201833880} {\bibfield
  {journal} {\bibinfo  {journal} {\aap}\ }\textbf {\bibinfo {volume} {641}},\
  \bibinfo {eid} {A1} (\bibinfo {year} {2020})}\BibitemShut {NoStop}%
\bibitem [{\citenamefont {{Reischke}}\ \emph {et~al.}(2021)\citenamefont
  {{Reischke}}, \citenamefont {{Hagstotz}},\ and\ \citenamefont
  {{Lilow}}}]{Reischke_2021}%
  \BibitemOpen
  \bibfield  {author} {\bibinfo {author} {\bibfnamefont {R.}~\bibnamefont
  {{Reischke}}}, \bibinfo {author} {\bibfnamefont {S.}~\bibnamefont
  {{Hagstotz}}}, \ and\ \bibinfo {author} {\bibfnamefont {R.}~\bibnamefont
  {{Lilow}}},\ }\href@noop {} {\bibfield  {journal} {\bibinfo  {journal} {arXiv
  e-prints}\ ,\ \bibinfo {eid} {arXiv:2102.11554}} (\bibinfo {year}
  {2021})}\BibitemShut {NoStop}%
\bibitem [{\citenamefont {{Alonso}}(2021)}]{Alonso_2021}%
  \BibitemOpen
  \bibfield  {author} {\bibinfo {author} {\bibfnamefont {D.}~\bibnamefont
  {{Alonso}}},\ }\href {\doibase 10.1103/PhysRevD.103.123544} {\bibfield
  {journal} {\bibinfo  {journal} {\prd}\ }\textbf {\bibinfo {volume} {103}},\
  \bibinfo {eid} {123544} (\bibinfo {year} {2021})}\BibitemShut {NoStop}%
\bibitem [{\citenamefont {{Meiksin}}(2009)}]{Meiksin_2009}%
  \BibitemOpen
  \bibfield  {author} {\bibinfo {author} {\bibfnamefont {A.~A.}\ \bibnamefont
  {{Meiksin}}},\ }\href {\doibase 10.1103/RevModPhys.81.1405} {\bibfield
  {journal} {\bibinfo  {journal} {Reviews of Modern Physics}\ }\textbf
  {\bibinfo {volume} {81}},\ \bibinfo {pages} {1405} (\bibinfo {year}
  {2009})}\BibitemShut {NoStop}%
\bibitem [{\citenamefont {{Becker}}\ \emph {et~al.}(2011)\citenamefont
  {{Becker}}, \citenamefont {{Bolton}}, \citenamefont {{Haehnelt}},\ and\
  \citenamefont {{Sargent}}}]{Becker_2011}%
  \BibitemOpen
  \bibfield  {author} {\bibinfo {author} {\bibfnamefont {G.~D.}\ \bibnamefont
  {{Becker}}}, \bibinfo {author} {\bibfnamefont {J.~S.}\ \bibnamefont
  {{Bolton}}}, \bibinfo {author} {\bibfnamefont {M.~G.}\ \bibnamefont
  {{Haehnelt}}}, \ and\ \bibinfo {author} {\bibfnamefont {W.~L.~W.}\
  \bibnamefont {{Sargent}}},\ }\href {\doibase
  10.1111/j.1365-2966.2010.17507.x} {\bibfield  {journal} {\bibinfo  {journal}
  {\mnras}\ }\textbf {\bibinfo {volume} {410}},\ \bibinfo {pages} {1096}
  (\bibinfo {year} {2011})}\BibitemShut {NoStop}%
\bibitem [{\citenamefont {{Shull}}\ \emph {et~al.}(2012)\citenamefont
  {{Shull}}, \citenamefont {{Smith}},\ and\ \citenamefont
  {{Danforth}}}]{Shull_2012}%
  \BibitemOpen
  \bibfield  {author} {\bibinfo {author} {\bibfnamefont {J.~M.}\ \bibnamefont
  {{Shull}}}, \bibinfo {author} {\bibfnamefont {B.~D.}\ \bibnamefont
  {{Smith}}}, \ and\ \bibinfo {author} {\bibfnamefont {C.~W.}\ \bibnamefont
  {{Danforth}}},\ }\href {\doibase 10.1088/0004-637X/759/1/23} {\bibfield
  {journal} {\bibinfo  {journal} {\apj}\ }\textbf {\bibinfo {volume} {759}},\
  \bibinfo {eid} {23} (\bibinfo {year} {2012})}\BibitemShut {NoStop}%
\bibitem [{\citenamefont {{Hakkila}}\ \emph {et~al.}(2007)\citenamefont
  {{Hakkila}}, \citenamefont {{Giblin}}, \citenamefont {{Young}}, \citenamefont
  {{Fuller}}, \citenamefont {{Peters}}, \citenamefont {{Nolan}}, \citenamefont
  {{Sonnett}}, \citenamefont {{Haglin}},\ and\ \citenamefont
  {{Roiger}}}]{Hakkila_2007}%
  \BibitemOpen
  \bibfield  {author} {\bibinfo {author} {\bibfnamefont {J.}~\bibnamefont
  {{Hakkila}}}, \bibinfo {author} {\bibfnamefont {T.~W.}\ \bibnamefont
  {{Giblin}}}, \bibinfo {author} {\bibfnamefont {K.~C.}\ \bibnamefont
  {{Young}}}, \bibinfo {author} {\bibfnamefont {S.~P.}\ \bibnamefont
  {{Fuller}}}, \bibinfo {author} {\bibfnamefont {C.~D.}\ \bibnamefont
  {{Peters}}}, \bibinfo {author} {\bibfnamefont {C.}~\bibnamefont {{Nolan}}},
  \bibinfo {author} {\bibfnamefont {S.~M.}\ \bibnamefont {{Sonnett}}}, \bibinfo
  {author} {\bibfnamefont {D.~J.}\ \bibnamefont {{Haglin}}}, \ and\ \bibinfo
  {author} {\bibfnamefont {R.~J.}\ \bibnamefont {{Roiger}}},\ }\href {\doibase
  10.1086/511306} {\bibfield  {journal} {\bibinfo  {journal} {\apjs}\ }\textbf
  {\bibinfo {volume} {169}},\ \bibinfo {pages} {62} (\bibinfo {year}
  {2007})}\BibitemShut {NoStop}%
\bibitem [{\citenamefont {{Yu}}\ \emph {et~al.}(2018)\citenamefont {{Yu}},
  \citenamefont {{Xi}},\ and\ \citenamefont {{Wang}}}]{Yu_2018}%
  \BibitemOpen
  \bibfield  {author} {\bibinfo {author} {\bibfnamefont {H.}~\bibnamefont
  {{Yu}}}, \bibinfo {author} {\bibfnamefont {S.-Q.}\ \bibnamefont {{Xi}}}, \
  and\ \bibinfo {author} {\bibfnamefont {F.-Y.}\ \bibnamefont {{Wang}}},\
  }\href {\doibase 10.3847/1538-4357/aac2e3} {\bibfield  {journal} {\bibinfo
  {journal} {\apj}\ }\textbf {\bibinfo {volume} {860}},\ \bibinfo {eid} {173}
  (\bibinfo {year} {2018})}\BibitemShut {NoStop}%
\bibitem [{\citenamefont {{Yonetoku}}\ \emph {et~al.}(2004)\citenamefont
  {{Yonetoku}}, \citenamefont {{Murakami}}, \citenamefont {{Nakamura}},
  \citenamefont {{Yamazaki}}, \citenamefont {{Inoue}},\ and\ \citenamefont
  {{Ioka}}}]{Yonetoku_2004}%
  \BibitemOpen
  \bibfield  {author} {\bibinfo {author} {\bibfnamefont {D.}~\bibnamefont
  {{Yonetoku}}}, \bibinfo {author} {\bibfnamefont {T.}~\bibnamefont
  {{Murakami}}}, \bibinfo {author} {\bibfnamefont {T.}~\bibnamefont
  {{Nakamura}}}, \bibinfo {author} {\bibfnamefont {R.}~\bibnamefont
  {{Yamazaki}}}, \bibinfo {author} {\bibfnamefont {A.~K.}\ \bibnamefont
  {{Inoue}}}, \ and\ \bibinfo {author} {\bibfnamefont {K.}~\bibnamefont
  {{Ioka}}},\ }\href {\doibase 10.1086/421285} {\bibfield  {journal} {\bibinfo
  {journal} {\apj}\ }\textbf {\bibinfo {volume} {609}},\ \bibinfo {pages} {935}
  (\bibinfo {year} {2004})}\BibitemShut {NoStop}%
\bibitem [{\citenamefont {{Shen}}\ \emph {et~al.}(2005)\citenamefont {{Shen}},
  \citenamefont {{Song}},\ and\ \citenamefont {{Li}}}]{Shen_2005}%
  \BibitemOpen
  \bibfield  {author} {\bibinfo {author} {\bibfnamefont {R.-F.}\ \bibnamefont
  {{Shen}}}, \bibinfo {author} {\bibfnamefont {L.-M.}\ \bibnamefont {{Song}}},
  \ and\ \bibinfo {author} {\bibfnamefont {Z.}~\bibnamefont {{Li}}},\ }\href
  {\doibase 10.1111/j.1365-2966.2005.09163.x} {\bibfield  {journal} {\bibinfo
  {journal} {\mnras}\ }\textbf {\bibinfo {volume} {362}},\ \bibinfo {pages}
  {59} (\bibinfo {year} {2005})}\BibitemShut {NoStop}%
\bibitem [{\citenamefont {{Lu}}\ \emph {et~al.}(2006)\citenamefont {{Lu}},
  \citenamefont {{Qin}}, \citenamefont {{Zhang}},\ and\ \citenamefont
  {{Yi}}}]{Lu_2006}%
  \BibitemOpen
  \bibfield  {author} {\bibinfo {author} {\bibfnamefont {R.~J.}\ \bibnamefont
  {{Lu}}}, \bibinfo {author} {\bibfnamefont {Y.~P.}\ \bibnamefont {{Qin}}},
  \bibinfo {author} {\bibfnamefont {Z.~B.}\ \bibnamefont {{Zhang}}}, \ and\
  \bibinfo {author} {\bibfnamefont {T.~F.}\ \bibnamefont {{Yi}}},\ }\href
  {\doibase 10.1111/j.1365-2966.2005.09951.x} {\bibfield  {journal} {\bibinfo
  {journal} {\mnras}\ }\textbf {\bibinfo {volume} {367}},\ \bibinfo {pages}
  {275} (\bibinfo {year} {2006})}\BibitemShut {NoStop}%
\bibitem [{\citenamefont {{Daigne}}\ and\ \citenamefont
  {{Mochkovitch}}(2003)}]{Daigne_2003}%
  \BibitemOpen
  \bibfield  {author} {\bibinfo {author} {\bibfnamefont {F.}~\bibnamefont
  {{Daigne}}}\ and\ \bibinfo {author} {\bibfnamefont {R.}~\bibnamefont
  {{Mochkovitch}}},\ }\href {\doibase 10.1046/j.1365-8711.2003.06575.x}
  {\bibfield  {journal} {\bibinfo  {journal} {\mnras}\ }\textbf {\bibinfo
  {volume} {342}},\ \bibinfo {pages} {587} (\bibinfo {year}
  {2003})}\BibitemShut {NoStop}%
\bibitem [{\citenamefont {{Peng}}\ \emph {et~al.}(2011)\citenamefont {{Peng}},
  \citenamefont {{Yin}}, \citenamefont {{Bi}}, \citenamefont {{Bao}},\ and\
  \citenamefont {{Ma}}}]{Peng_2011}%
  \BibitemOpen
  \bibfield  {author} {\bibinfo {author} {\bibfnamefont {Z.~Y.}\ \bibnamefont
  {{Peng}}}, \bibinfo {author} {\bibfnamefont {Y.}~\bibnamefont {{Yin}}},
  \bibinfo {author} {\bibfnamefont {X.~W.}\ \bibnamefont {{Bi}}}, \bibinfo
  {author} {\bibfnamefont {Y.~Y.}\ \bibnamefont {{Bao}}}, \ and\ \bibinfo
  {author} {\bibfnamefont {L.}~\bibnamefont {{Ma}}},\ }\href {\doibase
  10.1002/asna.201011474} {\bibfield  {journal} {\bibinfo  {journal}
  {Astronomische Nachrichten}\ }\textbf {\bibinfo {volume} {332}},\ \bibinfo
  {pages} {92} (\bibinfo {year} {2011})}\BibitemShut {NoStop}%
\bibitem [{\citenamefont {{Du}}\ \emph {et~al.}(2019)\citenamefont {{Du}},
  \citenamefont {{Lin}}, \citenamefont {{Lu}}, \citenamefont {{Li}},
  \citenamefont {{Gan}}, \citenamefont {{Ren}}, \citenamefont {{Xiang-Gao}},\
  and\ \citenamefont {{Liang}}}]{Du_2019}%
  \BibitemOpen
  \bibfield  {author} {\bibinfo {author} {\bibfnamefont {S.-S.}\ \bibnamefont
  {{Du}}}, \bibinfo {author} {\bibfnamefont {D.-B.}\ \bibnamefont {{Lin}}},
  \bibinfo {author} {\bibfnamefont {R.-J.}\ \bibnamefont {{Lu}}}, \bibinfo
  {author} {\bibfnamefont {R.-Q.}\ \bibnamefont {{Li}}}, \bibinfo {author}
  {\bibfnamefont {Y.-Y.}\ \bibnamefont {{Gan}}}, \bibinfo {author}
  {\bibfnamefont {J.}~\bibnamefont {{Ren}}}, \bibinfo {author} {\bibfnamefont
  {W.}~\bibnamefont {{Xiang-Gao}}}, \ and\ \bibinfo {author} {\bibfnamefont
  {E.-W.}\ \bibnamefont {{Liang}}},\ }\href {\doibase 10.3847/1538-4357/ab33fe}
  {\bibfield  {journal} {\bibinfo  {journal} {\apj}\ }\textbf {\bibinfo
  {volume} {882}},\ \bibinfo {eid} {115} (\bibinfo {year} {2019})}\BibitemShut
  {NoStop}%
\bibitem [{\citenamefont {{Lu}}\ \emph {et~al.}(2018)\citenamefont {{Lu}},
  \citenamefont {{Liang}}, \citenamefont {{Lin}}, \citenamefont {{L{\"u}}},
  \citenamefont {{Wang}}, \citenamefont {{L{\"u}}}, \citenamefont {{Liu}},
  \citenamefont {{Liang}},\ and\ \citenamefont {{Zhang}}}]{Lu_2018}%
  \BibitemOpen
  \bibfield  {author} {\bibinfo {author} {\bibfnamefont {R.-J.}\ \bibnamefont
  {{Lu}}}, \bibinfo {author} {\bibfnamefont {Y.-F.}\ \bibnamefont {{Liang}}},
  \bibinfo {author} {\bibfnamefont {D.-B.}\ \bibnamefont {{Lin}}}, \bibinfo
  {author} {\bibfnamefont {J.}~\bibnamefont {{L{\"u}}}}, \bibinfo {author}
  {\bibfnamefont {X.-G.}\ \bibnamefont {{Wang}}}, \bibinfo {author}
  {\bibfnamefont {H.-J.}\ \bibnamefont {{L{\"u}}}}, \bibinfo {author}
  {\bibfnamefont {H.-B.}\ \bibnamefont {{Liu}}}, \bibinfo {author}
  {\bibfnamefont {E.-W.}\ \bibnamefont {{Liang}}}, \ and\ \bibinfo {author}
  {\bibfnamefont {B.}~\bibnamefont {{Zhang}}},\ }\href {\doibase
  10.3847/1538-4357/aada16} {\bibfield  {journal} {\bibinfo  {journal} {\apj}\
  }\textbf {\bibinfo {volume} {865}},\ \bibinfo {eid} {153} (\bibinfo {year}
  {2018})}\BibitemShut {NoStop}%
\bibitem [{\citenamefont {{Uhm}}\ and\ \citenamefont
  {{Zhang}}(2016)}]{Uhm_2016}%
  \BibitemOpen
  \bibfield  {author} {\bibinfo {author} {\bibfnamefont {Z.~L.}\ \bibnamefont
  {{Uhm}}}\ and\ \bibinfo {author} {\bibfnamefont {B.}~\bibnamefont
  {{Zhang}}},\ }\href {\doibase 10.3847/0004-637X/825/2/97} {\bibfield
  {journal} {\bibinfo  {journal} {\apj}\ }\textbf {\bibinfo {volume} {825}},\
  \bibinfo {eid} {97} (\bibinfo {year} {2016})}\BibitemShut {NoStop}%
\bibitem [{\citenamefont {Bartlett}\ \emph {et~al.}(2021)\citenamefont
  {Bartlett}, \citenamefont {Desmond},\ and\ \citenamefont
  {Ferreira}}]{Bartlett_2021_HAGN}%
  \BibitemOpen
  \bibfield  {author} {\bibinfo {author} {\bibfnamefont {D.~J.}\ \bibnamefont
  {Bartlett}}, \bibinfo {author} {\bibfnamefont {H.}~\bibnamefont {Desmond}}, \
  and\ \bibinfo {author} {\bibfnamefont {P.~G.}\ \bibnamefont {Ferreira}},\
  }\href {\doibase 10.1103/PhysRevD.103.123502} {\bibfield  {journal} {\bibinfo
   {journal} {Phys. Rev. D}\ }\textbf {\bibinfo {volume} {103}},\ \bibinfo
  {pages} {123502} (\bibinfo {year} {2021})}\BibitemShut {NoStop}%
\bibitem [{\citenamefont {{Moutarde}}\ \emph {et~al.}(1991)\citenamefont
  {{Moutarde}}, \citenamefont {{Alimi}}, \citenamefont {{Bouchet}},
  \citenamefont {{Pellat}},\ and\ \citenamefont {{Ramani}}}]{MOUTARDE1991}%
  \BibitemOpen
  \bibfield  {author} {\bibinfo {author} {\bibfnamefont {F.}~\bibnamefont
  {{Moutarde}}}, \bibinfo {author} {\bibfnamefont {J.}~\bibnamefont {{Alimi}}},
  \bibinfo {author} {\bibfnamefont {F.~R.}\ \bibnamefont {{Bouchet}}}, \bibinfo
  {author} {\bibfnamefont {R.}~\bibnamefont {{Pellat}}}, \ and\ \bibinfo
  {author} {\bibfnamefont {A.}~\bibnamefont {{Ramani}}},\ }\href {\doibase
  10.1086/170728} {\bibfield  {journal} {\bibinfo  {journal} {\apj}\ }\textbf
  {\bibinfo {volume} {382}},\ \bibinfo {pages} {377} (\bibinfo {year}
  {1991})}\BibitemShut {NoStop}%
\bibitem [{\citenamefont {{Buchert}}\ \emph {et~al.}(1994)\citenamefont
  {{Buchert}}, \citenamefont {{Melott}},\ and\ \citenamefont
  {{Weiss}}}]{BUCHERT1994}%
  \BibitemOpen
  \bibfield  {author} {\bibinfo {author} {\bibfnamefont {T.}~\bibnamefont
  {{Buchert}}}, \bibinfo {author} {\bibfnamefont {A.~L.}\ \bibnamefont
  {{Melott}}}, \ and\ \bibinfo {author} {\bibfnamefont {A.~G.}\ \bibnamefont
  {{Weiss}}},\ }\href@noop {} {\bibfield  {journal} {\bibinfo  {journal}
  {\aap}\ }\textbf {\bibinfo {volume} {288}},\ \bibinfo {pages} {349} (\bibinfo
  {year} {1994})}\BibitemShut {NoStop}%
\bibitem [{\citenamefont {{Bouchet}}\ \emph {et~al.}(1995)\citenamefont
  {{Bouchet}}, \citenamefont {{Colombi}}, \citenamefont {{Hivon}},\ and\
  \citenamefont {{Juszkiewicz}}}]{BOUCHET1995}%
  \BibitemOpen
  \bibfield  {author} {\bibinfo {author} {\bibfnamefont {F.~R.}\ \bibnamefont
  {{Bouchet}}}, \bibinfo {author} {\bibfnamefont {S.}~\bibnamefont
  {{Colombi}}}, \bibinfo {author} {\bibfnamefont {E.}~\bibnamefont {{Hivon}}},
  \ and\ \bibinfo {author} {\bibfnamefont {R.}~\bibnamefont {{Juszkiewicz}}},\
  }\href@noop {} {\bibfield  {journal} {\bibinfo  {journal} {\aap}\ }\textbf
  {\bibinfo {volume} {296}},\ \bibinfo {pages} {575} (\bibinfo {year}
  {1995})}\BibitemShut {NoStop}%
\bibitem [{\citenamefont {{Scoccimarro}}(2000)}]{SCOCCIMARRO2000}%
  \BibitemOpen
  \bibfield  {author} {\bibinfo {author} {\bibfnamefont {R.}~\bibnamefont
  {{Scoccimarro}}},\ }\href {\doibase 10.1086/317248} {\bibfield  {journal}
  {\bibinfo  {journal} {\apj}\ }\textbf {\bibinfo {volume} {544}},\ \bibinfo
  {pages} {597} (\bibinfo {year} {2000})}\BibitemShut {NoStop}%
\bibitem [{\citenamefont {{Leclercq}}\ \emph {et~al.}(2013)\citenamefont
  {{Leclercq}}, \citenamefont {{Jasche}}, \citenamefont {{Gil-Mar{\'{\i}}n}},\
  and\ \citenamefont {{Wandelt}}}]{Leclercq_2013}%
  \BibitemOpen
  \bibfield  {author} {\bibinfo {author} {\bibfnamefont {F.}~\bibnamefont
  {{Leclercq}}}, \bibinfo {author} {\bibfnamefont {J.}~\bibnamefont
  {{Jasche}}}, \bibinfo {author} {\bibfnamefont {H.}~\bibnamefont
  {{Gil-Mar{\'{\i}}n}}}, \ and\ \bibinfo {author} {\bibfnamefont
  {B.}~\bibnamefont {{Wandelt}}},\ }\href {\doibase
  10.1088/1475-7516/2013/11/048} {\bibfield  {journal} {\bibinfo  {journal}
  {\jcap}\ }\textbf {\bibinfo {volume} {11}},\ \bibinfo {eid} {048} (\bibinfo
  {year} {2013})}\BibitemShut {NoStop}%
\bibitem [{\citenamefont {{Eisenstein}}\ \emph {et~al.}(2011)\citenamefont
  {{Eisenstein}} \emph {et~al.}}]{Eisenstein_2011}%
  \BibitemOpen
  \bibfield  {author} {\bibinfo {author} {\bibfnamefont {D.~J.}\ \bibnamefont
  {{Eisenstein}}} \emph {et~al.},\ }\href {\doibase 10.1088/0004-6256/142/3/72}
  {\bibfield  {journal} {\bibinfo  {journal} {\aj}\ }\textbf {\bibinfo {volume}
  {142}},\ \bibinfo {eid} {72} (\bibinfo {year} {2011})}\BibitemShut {NoStop}%
\bibitem [{\citenamefont {{Lavaux}}\ \emph {et~al.}(2019)\citenamefont
  {{Lavaux}}, \citenamefont {{Jasche}},\ and\ \citenamefont
  {{Leclercq}}}]{Lavaux_2019}%
  \BibitemOpen
  \bibfield  {author} {\bibinfo {author} {\bibfnamefont {G.}~\bibnamefont
  {{Lavaux}}}, \bibinfo {author} {\bibfnamefont {J.}~\bibnamefont {{Jasche}}},
  \ and\ \bibinfo {author} {\bibfnamefont {F.}~\bibnamefont {{Leclercq}}},\
  }\href@noop {} {\bibfield  {journal} {\bibinfo  {journal} {arXiv e-prints}\
  ,\ \bibinfo {eid} {arXiv:1909.06396}} (\bibinfo {year} {2019})}\BibitemShut
  {NoStop}%
\bibitem [{\citenamefont {Jasche}\ and\ \citenamefont
  {Lavaux}(2019)}]{Jasche_Lavaux}%
  \BibitemOpen
  \bibfield  {author} {\bibinfo {author} {\bibfnamefont {J.}~\bibnamefont
  {Jasche}}\ and\ \bibinfo {author} {\bibfnamefont {G.}~\bibnamefont
  {Lavaux}},\ }\href {\doibase 10.1051/0004-6361/201833710} {\bibfield
  {journal} {\bibinfo  {journal} {Astron. Astrophys.}\ }\textbf {\bibinfo
  {volume} {625}},\ \bibinfo {pages} {A64} (\bibinfo {year}
  {2019})}\BibitemShut {NoStop}%
\bibitem [{\citenamefont {{Bartlett}}\ \emph {et~al.}(2021)\citenamefont
  {{Bartlett}}, \citenamefont {{Desmond}},\ and\ \citenamefont
  {{Ferreira}}}]{Bartlett_2021_VS}%
  \BibitemOpen
  \bibfield  {author} {\bibinfo {author} {\bibfnamefont {D.~J.}\ \bibnamefont
  {{Bartlett}}}, \bibinfo {author} {\bibfnamefont {H.}~\bibnamefont
  {{Desmond}}}, \ and\ \bibinfo {author} {\bibfnamefont {P.~G.}\ \bibnamefont
  {{Ferreira}}},\ }\href {\doibase 10.1103/PhysRevD.103.023523} {\bibfield
  {journal} {\bibinfo  {journal} {\prd}\ }\textbf {\bibinfo {volume} {103}},\
  \bibinfo {eid} {023523} (\bibinfo {year} {2021})}\BibitemShut {NoStop}%
\bibitem [{\citenamefont {{Chisari}}\ \emph {et~al.}(2019)\citenamefont
  {{Chisari}} \emph {et~al.}}]{CCL_2019}%
  \BibitemOpen
  \bibfield  {author} {\bibinfo {author} {\bibfnamefont {N.~E.}\ \bibnamefont
  {{Chisari}}} \emph {et~al.},\ }\href {\doibase 10.3847/1538-4365/ab1658}
  {\bibfield  {journal} {\bibinfo  {journal} {\apjs}\ }\textbf {\bibinfo
  {volume} {242}},\ \bibinfo {eid} {2} (\bibinfo {year} {2019})}\BibitemShut
  {NoStop}%
\bibitem [{\citenamefont {{Lemos}}\ \emph {et~al.}(2017)\citenamefont
  {{Lemos}}, \citenamefont {{Challinor}},\ and\ \citenamefont
  {{Efstathiou}}}]{Lemos_2017}%
  \BibitemOpen
  \bibfield  {author} {\bibinfo {author} {\bibfnamefont {P.}~\bibnamefont
  {{Lemos}}}, \bibinfo {author} {\bibfnamefont {A.}~\bibnamefont
  {{Challinor}}}, \ and\ \bibinfo {author} {\bibfnamefont {G.}~\bibnamefont
  {{Efstathiou}}},\ }\href {\doibase 10.1088/1475-7516/2017/05/014} {\bibfield
  {journal} {\bibinfo  {journal} {Journal of Cosmology and Astro-Particle
  Physics}\ }\textbf {\bibinfo {volume} {2017}},\ \bibinfo {eid} {014}
  (\bibinfo {year} {2017})}\BibitemShut {NoStop}%
\bibitem [{\citenamefont {Zonca}\ \emph {et~al.}(2019)\citenamefont {Zonca},
  \citenamefont {Singer}, \citenamefont {Lenz}, \citenamefont {Reinecke},
  \citenamefont {Rosset}, \citenamefont {Hivon},\ and\ \citenamefont
  {Gorski}}]{Zonca_2019}%
  \BibitemOpen
  \bibfield  {author} {\bibinfo {author} {\bibfnamefont {A.}~\bibnamefont
  {Zonca}}, \bibinfo {author} {\bibfnamefont {L.}~\bibnamefont {Singer}},
  \bibinfo {author} {\bibfnamefont {D.}~\bibnamefont {Lenz}}, \bibinfo {author}
  {\bibfnamefont {M.}~\bibnamefont {Reinecke}}, \bibinfo {author}
  {\bibfnamefont {C.}~\bibnamefont {Rosset}}, \bibinfo {author} {\bibfnamefont
  {E.}~\bibnamefont {Hivon}}, \ and\ \bibinfo {author} {\bibfnamefont
  {K.}~\bibnamefont {Gorski}},\ }\href {\doibase 10.21105/joss.01298}
  {\bibfield  {journal} {\bibinfo  {journal} {Journal of Open Source Software}\
  }\textbf {\bibinfo {volume} {4}},\ \bibinfo {pages} {1298} (\bibinfo {year}
  {2019})}\BibitemShut {NoStop}%
\bibitem [{\citenamefont {{G{\'o}rski}}\ \emph {et~al.}(2005)\citenamefont
  {{G{\'o}rski}}, \citenamefont {{Hivon}}, \citenamefont {{Banday}},
  \citenamefont {{Wandelt}}, \citenamefont {{Hansen}}, \citenamefont
  {{Reinecke}},\ and\ \citenamefont {{Bartelmann}}}]{Gorski_2005}%
  \BibitemOpen
  \bibfield  {author} {\bibinfo {author} {\bibfnamefont {K.~M.}\ \bibnamefont
  {{G{\'o}rski}}}, \bibinfo {author} {\bibfnamefont {E.}~\bibnamefont
  {{Hivon}}}, \bibinfo {author} {\bibfnamefont {A.~J.}\ \bibnamefont
  {{Banday}}}, \bibinfo {author} {\bibfnamefont {B.~D.}\ \bibnamefont
  {{Wandelt}}}, \bibinfo {author} {\bibfnamefont {F.~K.}\ \bibnamefont
  {{Hansen}}}, \bibinfo {author} {\bibfnamefont {M.}~\bibnamefont
  {{Reinecke}}}, \ and\ \bibinfo {author} {\bibfnamefont {M.}~\bibnamefont
  {{Bartelmann}}},\ }\href {\doibase 10.1086/427976} {\bibfield  {journal}
  {\bibinfo  {journal} {\apj}\ }\textbf {\bibinfo {volume} {622}},\ \bibinfo
  {pages} {759} (\bibinfo {year} {2005})}\BibitemShut {NoStop}%
\bibitem [{\citenamefont {Pedregosa}\ \emph {et~al.}(2011)\citenamefont
  {Pedregosa} \emph {et~al.}}]{scikit-learn}%
  \BibitemOpen
  \bibfield  {author} {\bibinfo {author} {\bibfnamefont {F.}~\bibnamefont
  {Pedregosa}} \emph {et~al.},\ }\href@noop {} {\bibfield  {journal} {\bibinfo
  {journal} {Journal of Machine Learning Research}\ }\textbf {\bibinfo {volume}
  {12}},\ \bibinfo {pages} {2825} (\bibinfo {year} {2011})}\BibitemShut
  {NoStop}%
\bibitem [{\citenamefont {{Feroz}}\ and\ \citenamefont
  {{Hobson}}(2008)}]{Multinest_1}%
  \BibitemOpen
  \bibfield  {author} {\bibinfo {author} {\bibfnamefont {F.}~\bibnamefont
  {{Feroz}}}\ and\ \bibinfo {author} {\bibfnamefont {M.~P.}\ \bibnamefont
  {{Hobson}}},\ }\href {\doibase 10.1111/j.1365-2966.2007.12353.x} {\bibfield
  {journal} {\bibinfo  {journal} {\mnras}\ }\textbf {\bibinfo {volume} {384}},\
  \bibinfo {pages} {449} (\bibinfo {year} {2008})}\BibitemShut {NoStop}%
\bibitem [{\citenamefont {{Feroz}}\ \emph {et~al.}(2009)\citenamefont
  {{Feroz}}, \citenamefont {{Hobson}},\ and\ \citenamefont
  {{Bridges}}}]{Multinest_2}%
  \BibitemOpen
  \bibfield  {author} {\bibinfo {author} {\bibfnamefont {F.}~\bibnamefont
  {{Feroz}}}, \bibinfo {author} {\bibfnamefont {M.~P.}\ \bibnamefont
  {{Hobson}}}, \ and\ \bibinfo {author} {\bibfnamefont {M.}~\bibnamefont
  {{Bridges}}},\ }\href {\doibase 10.1111/j.1365-2966.2009.14548.x} {\bibfield
  {journal} {\bibinfo  {journal} {\mnras}\ }\textbf {\bibinfo {volume} {398}},\
  \bibinfo {pages} {1601} (\bibinfo {year} {2009})}\BibitemShut {NoStop}%
\bibitem [{\citenamefont {{Feroz}}\ \emph {et~al.}(2019)\citenamefont
  {{Feroz}}, \citenamefont {{Hobson}}, \citenamefont {{Cameron}},\ and\
  \citenamefont {{Pettitt}}}]{Multinest_3}%
  \BibitemOpen
  \bibfield  {author} {\bibinfo {author} {\bibfnamefont {F.}~\bibnamefont
  {{Feroz}}}, \bibinfo {author} {\bibfnamefont {M.~P.}\ \bibnamefont
  {{Hobson}}}, \bibinfo {author} {\bibfnamefont {E.}~\bibnamefont {{Cameron}}},
  \ and\ \bibinfo {author} {\bibfnamefont {A.~N.}\ \bibnamefont {{Pettitt}}},\
  }\href {\doibase 10.21105/astro.1306.2144} {\bibfield  {journal} {\bibinfo
  {journal} {The Open Journal of Astrophysics}\ }\textbf {\bibinfo {volume}
  {2}},\ \bibinfo {eid} {10} (\bibinfo {year} {2019})}\BibitemShut {NoStop}%
\bibitem [{\citenamefont {{Handley}}\ and\ \citenamefont
  {{Lemos}}(2019)}]{Handley_2019}%
  \BibitemOpen
  \bibfield  {author} {\bibinfo {author} {\bibfnamefont {W.}~\bibnamefont
  {{Handley}}}\ and\ \bibinfo {author} {\bibfnamefont {P.}~\bibnamefont
  {{Lemos}}},\ }\href {\doibase 10.1103/PhysRevD.100.023512} {\bibfield
  {journal} {\bibinfo  {journal} {\prd}\ }\textbf {\bibinfo {volume} {100}},\
  \bibinfo {eid} {023512} (\bibinfo {year} {2019})}\BibitemShut {NoStop}%
\bibitem [{\citenamefont {Handley}(2019)}]{anesthetic}%
  \BibitemOpen
  \bibfield  {author} {\bibinfo {author} {\bibfnamefont {W.}~\bibnamefont
  {Handley}},\ }\href {\doibase 10.21105/joss.01414} {\bibfield  {journal}
  {\bibinfo  {journal} {The Journal of Open Source Software}\ }\textbf
  {\bibinfo {volume} {4}},\ \bibinfo {pages} {1414} (\bibinfo {year}
  {2019})}\BibitemShut {NoStop}%
\bibitem [{\citenamefont {{Kocevski}}\ and\ \citenamefont
  {{Liang}}(2003)}]{Kocevski_2003}%
  \BibitemOpen
  \bibfield  {author} {\bibinfo {author} {\bibfnamefont {D.}~\bibnamefont
  {{Kocevski}}}\ and\ \bibinfo {author} {\bibfnamefont {E.}~\bibnamefont
  {{Liang}}},\ }\href {\doibase 10.1086/376868} {\bibfield  {journal} {\bibinfo
   {journal} {\apj}\ }\textbf {\bibinfo {volume} {594}},\ \bibinfo {pages}
  {385} (\bibinfo {year} {2003})}\BibitemShut {NoStop}%
\bibitem [{\citenamefont {{Schaefer}}(2004)}]{Schaefer_2004}%
  \BibitemOpen
  \bibfield  {author} {\bibinfo {author} {\bibfnamefont {B.~E.}\ \bibnamefont
  {{Schaefer}}},\ }\href {\doibase 10.1086/380898} {\bibfield  {journal}
  {\bibinfo  {journal} {\apj}\ }\textbf {\bibinfo {volume} {602}},\ \bibinfo
  {pages} {306} (\bibinfo {year} {2004})}\BibitemShut {NoStop}%
\bibitem [{\citenamefont {{Ryde}}(2005)}]{Ryde_2005}%
  \BibitemOpen
  \bibfield  {author} {\bibinfo {author} {\bibfnamefont {F.}~\bibnamefont
  {{Ryde}}},\ }\href {\doibase 10.1051/0004-6361:20041364} {\bibfield
  {journal} {\bibinfo  {journal} {\aap}\ }\textbf {\bibinfo {volume} {429}},\
  \bibinfo {pages} {869} (\bibinfo {year} {2005})}\BibitemShut {NoStop}%
\bibitem [{\citenamefont {{Hakkila}}\ and\ \citenamefont
  {{Preece}}(2011)}]{Hakkila_2011}%
  \BibitemOpen
  \bibfield  {author} {\bibinfo {author} {\bibfnamefont {J.}~\bibnamefont
  {{Hakkila}}}\ and\ \bibinfo {author} {\bibfnamefont {R.~D.}\ \bibnamefont
  {{Preece}}},\ }\href {\doibase 10.1088/0004-637X/740/2/104} {\bibfield
  {journal} {\bibinfo  {journal} {\apj}\ }\textbf {\bibinfo {volume} {740}},\
  \bibinfo {eid} {104} (\bibinfo {year} {2011})}\BibitemShut {NoStop}%
\bibitem [{\citenamefont {{Sivaram}}(1999)}]{Sivaram_1999}%
  \BibitemOpen
  \bibfield  {author} {\bibinfo {author} {\bibfnamefont {C.}~\bibnamefont
  {{Sivaram}}},\ }\href@noop {} {\bibfield  {journal} {\bibinfo  {journal}
  {Bulletin of the Astronomical Society of India}\ }\textbf {\bibinfo {volume}
  {27}},\ \bibinfo {pages} {627} (\bibinfo {year} {1999})}\BibitemShut
  {NoStop}%
\bibitem [{\citenamefont {{Gao}}\ \emph {et~al.}(2015)\citenamefont {{Gao}},
  \citenamefont {{Wu}},\ and\ \citenamefont {{M{\'e}sz{\'a}ros}}}]{Gao_2015}%
  \BibitemOpen
  \bibfield  {author} {\bibinfo {author} {\bibfnamefont {H.}~\bibnamefont
  {{Gao}}}, \bibinfo {author} {\bibfnamefont {X.-F.}\ \bibnamefont {{Wu}}}, \
  and\ \bibinfo {author} {\bibfnamefont {P.}~\bibnamefont
  {{M{\'e}sz{\'a}ros}}},\ }\href {\doibase 10.1088/0004-637X/810/2/121}
  {\bibfield  {journal} {\bibinfo  {journal} {\apj}\ }\textbf {\bibinfo
  {volume} {810}},\ \bibinfo {eid} {121} (\bibinfo {year} {2015})}\BibitemShut
  {NoStop}%
\bibitem [{\citenamefont {{Sang}}\ \emph {et~al.}(2016)\citenamefont {{Sang}},
  \citenamefont {{Lin}},\ and\ \citenamefont {{Chang}}}]{Sang_2016}%
  \BibitemOpen
  \bibfield  {author} {\bibinfo {author} {\bibfnamefont {Y.}~\bibnamefont
  {{Sang}}}, \bibinfo {author} {\bibfnamefont {H.-N.}\ \bibnamefont {{Lin}}}, \
  and\ \bibinfo {author} {\bibfnamefont {Z.}~\bibnamefont {{Chang}}},\ }\href
  {\doibase 10.1093/mnras/stw1136} {\bibfield  {journal} {\bibinfo  {journal}
  {\mnras}\ }\textbf {\bibinfo {volume} {460}},\ \bibinfo {pages} {2282}
  (\bibinfo {year} {2016})}\BibitemShut {NoStop}%
\bibitem [{\citenamefont {Wei}\ \emph {et~al.}(2016)\citenamefont {Wei},
  \citenamefont {Wu}, \citenamefont {Gao},\ and\ \citenamefont
  {M{\'{e}}sz{\'{a}}ros}}]{Wei_2016}%
  \BibitemOpen
  \bibfield  {author} {\bibinfo {author} {\bibfnamefont {J.-J.}\ \bibnamefont
  {Wei}}, \bibinfo {author} {\bibfnamefont {X.-F.}\ \bibnamefont {Wu}},
  \bibinfo {author} {\bibfnamefont {H.}~\bibnamefont {Gao}}, \ and\ \bibinfo
  {author} {\bibfnamefont {P.}~\bibnamefont {M{\'{e}}sz{\'{a}}ros}},\ }\href
  {\doibase 10.1088/1475-7516/2016/08/031} {\bibfield  {journal} {\bibinfo
  {journal} {Journal of Cosmology and Astroparticle Physics}\ }\textbf
  {\bibinfo {volume} {2016}},\ \bibinfo {pages} {031} (\bibinfo {year}
  {2016})}\BibitemShut {NoStop}%
\bibitem [{\citenamefont {{Wei}}\ \emph {et~al.}(2015)\citenamefont {{Wei}},
  \citenamefont {{Gao}}, \citenamefont {{Wu}},\ and\ \citenamefont
  {{M{\'e}sz{\'a}ros}}}]{Wei_2015}%
  \BibitemOpen
  \bibfield  {author} {\bibinfo {author} {\bibfnamefont {J.-J.}\ \bibnamefont
  {{Wei}}}, \bibinfo {author} {\bibfnamefont {H.}~\bibnamefont {{Gao}}},
  \bibinfo {author} {\bibfnamefont {X.-F.}\ \bibnamefont {{Wu}}}, \ and\
  \bibinfo {author} {\bibfnamefont {P.}~\bibnamefont {{M{\'e}sz{\'a}ros}}},\
  }\href {\doibase 10.1103/PhysRevLett.115.261101} {\bibfield  {journal}
  {\bibinfo  {journal} {\prl}\ }\textbf {\bibinfo {volume} {115}},\ \bibinfo
  {eid} {261101} (\bibinfo {year} {2015})}\BibitemShut {NoStop}%
\bibitem [{\citenamefont {{Tingay}}\ and\ \citenamefont
  {{Kaplan}}(2016)}]{Tingay_2016}%
  \BibitemOpen
  \bibfield  {author} {\bibinfo {author} {\bibfnamefont {S.~J.}\ \bibnamefont
  {{Tingay}}}\ and\ \bibinfo {author} {\bibfnamefont {D.~L.}\ \bibnamefont
  {{Kaplan}}},\ }\href {\doibase 10.3847/2041-8205/820/2/L31} {\bibfield
  {journal} {\bibinfo  {journal} {\apjl}\ }\textbf {\bibinfo {volume} {820}},\
  \bibinfo {eid} {L31} (\bibinfo {year} {2016})}\BibitemShut {NoStop}%
\bibitem [{\citenamefont {{Xing}}\ \emph {et~al.}(2019)\citenamefont {{Xing}},
  \citenamefont {{Gao}}, \citenamefont {{Wei}}, \citenamefont {{Li}},
  \citenamefont {{Wang}}, \citenamefont {{Zhang}}, \citenamefont {{Wu}},\ and\
  \citenamefont {{M{\'e}sz{\'a}ros}}}]{Xing_2019}%
  \BibitemOpen
  \bibfield  {author} {\bibinfo {author} {\bibfnamefont {N.}~\bibnamefont
  {{Xing}}}, \bibinfo {author} {\bibfnamefont {H.}~\bibnamefont {{Gao}}},
  \bibinfo {author} {\bibfnamefont {J.-J.}\ \bibnamefont {{Wei}}}, \bibinfo
  {author} {\bibfnamefont {Z.}~\bibnamefont {{Li}}}, \bibinfo {author}
  {\bibfnamefont {W.}~\bibnamefont {{Wang}}}, \bibinfo {author} {\bibfnamefont
  {B.}~\bibnamefont {{Zhang}}}, \bibinfo {author} {\bibfnamefont {X.-F.}\
  \bibnamefont {{Wu}}}, \ and\ \bibinfo {author} {\bibfnamefont
  {P.}~\bibnamefont {{M{\'e}sz{\'a}ros}}},\ }\href {\doibase
  10.3847/2041-8213/ab3c5f} {\bibfield  {journal} {\bibinfo  {journal} {\apjl}\
  }\textbf {\bibinfo {volume} {882}},\ \bibinfo {eid} {L13} (\bibinfo {year}
  {2019})}\BibitemShut {NoStop}%
\bibitem [{\citenamefont {{Wang}}\ \emph {et~al.}(2020)\citenamefont {{Wang}},
  \citenamefont {{Li}},\ and\ \citenamefont {{Zhang}}}]{Wang_2020}%
  \BibitemOpen
  \bibfield  {author} {\bibinfo {author} {\bibfnamefont {D.}~\bibnamefont
  {{Wang}}}, \bibinfo {author} {\bibfnamefont {Z.}~\bibnamefont {{Li}}}, \ and\
  \bibinfo {author} {\bibfnamefont {J.}~\bibnamefont {{Zhang}}},\ }\href
  {\doibase 10.1016/j.dark.2020.100571} {\bibfield  {journal} {\bibinfo
  {journal} {Physics of the Dark Universe}\ }\textbf {\bibinfo {volume} {29}},\
  \bibinfo {eid} {100571} (\bibinfo {year} {2020})}\BibitemShut {NoStop}%
\bibitem [{\citenamefont {Longo}(1988)}]{Longo_1988}%
  \BibitemOpen
  \bibfield  {author} {\bibinfo {author} {\bibfnamefont {M.~J.}\ \bibnamefont
  {Longo}},\ }\href {\doibase 10.1103/PhysRevLett.60.173} {\bibfield  {journal}
  {\bibinfo  {journal} {Phys. Rev. Lett.}\ }\textbf {\bibinfo {volume} {60}},\
  \bibinfo {pages} {173} (\bibinfo {year} {1988})}\BibitemShut {NoStop}%
\bibitem [{\citenamefont {Krauss}\ and\ \citenamefont
  {Tremaine}(1988)}]{Krauss_1988}%
  \BibitemOpen
  \bibfield  {author} {\bibinfo {author} {\bibfnamefont {L.~M.}\ \bibnamefont
  {Krauss}}\ and\ \bibinfo {author} {\bibfnamefont {S.}~\bibnamefont
  {Tremaine}},\ }\href {\doibase 10.1103/PhysRevLett.60.176} {\bibfield
  {journal} {\bibinfo  {journal} {Phys. Rev. Lett.}\ }\textbf {\bibinfo
  {volume} {60}},\ \bibinfo {pages} {176} (\bibinfo {year} {1988})}\BibitemShut
  {NoStop}%
\bibitem [{\citenamefont {Wu}\ \emph {et~al.}(2016)\citenamefont {Wu},
  \citenamefont {Gao}, \citenamefont {Wei}, \citenamefont {M\'esz\'aros},
  \citenamefont {Zhang}, \citenamefont {Dai}, \citenamefont {Zhang},\ and\
  \citenamefont {Zhu}}]{Wu_2016}%
  \BibitemOpen
  \bibfield  {author} {\bibinfo {author} {\bibfnamefont {X.-F.}\ \bibnamefont
  {Wu}}, \bibinfo {author} {\bibfnamefont {H.}~\bibnamefont {Gao}}, \bibinfo
  {author} {\bibfnamefont {J.-J.}\ \bibnamefont {Wei}}, \bibinfo {author}
  {\bibfnamefont {P.}~\bibnamefont {M\'esz\'aros}}, \bibinfo {author}
  {\bibfnamefont {B.}~\bibnamefont {Zhang}}, \bibinfo {author} {\bibfnamefont
  {Z.-G.}\ \bibnamefont {Dai}}, \bibinfo {author} {\bibfnamefont {S.-N.}\
  \bibnamefont {Zhang}}, \ and\ \bibinfo {author} {\bibfnamefont {Z.-H.}\
  \bibnamefont {Zhu}},\ }\href {\doibase 10.1103/PhysRevD.94.024061} {\bibfield
   {journal} {\bibinfo  {journal} {Phys. Rev. D}\ }\textbf {\bibinfo {volume}
  {94}},\ \bibinfo {pages} {024061} (\bibinfo {year} {2016})}\BibitemShut
  {NoStop}%
\bibitem [{\citenamefont {{Kahya}}\ and\ \citenamefont
  {{Desai}}(2016)}]{Kahya_2016}%
  \BibitemOpen
  \bibfield  {author} {\bibinfo {author} {\bibfnamefont {E.~O.}\ \bibnamefont
  {{Kahya}}}\ and\ \bibinfo {author} {\bibfnamefont {S.}~\bibnamefont
  {{Desai}}},\ }\href {\doibase 10.1016/j.physletb.2016.03.033} {\bibfield
  {journal} {\bibinfo  {journal} {Physics Letters B}\ }\textbf {\bibinfo
  {volume} {756}},\ \bibinfo {pages} {265} (\bibinfo {year}
  {2016})}\BibitemShut {NoStop}%
\bibitem [{\citenamefont {{LIGO Scientific Collaboration and Virgo
  Collaboration}}(2017)}]{LIGO_2017}%
  \BibitemOpen
  \bibfield  {author} {\bibinfo {author} {\bibfnamefont {I.}~\bibnamefont
  {{LIGO Scientific Collaboration and Virgo Collaboration}}, \bibfnamefont
  {{Fermi Gamma-ray Burst Monitor}}},\ }\href {\doibase
  10.3847/2041-8213/aa920c} {\bibfield  {journal} {\bibinfo  {journal} {\apjl}\
  }\textbf {\bibinfo {volume} {848}},\ \bibinfo {eid} {L13} (\bibinfo {year}
  {2017})}\BibitemShut {NoStop}%
\bibitem [{\citenamefont {{Liu}}\ \emph {et~al.}(2017)\citenamefont {{Liu}},
  \citenamefont {{Zhao}}, \citenamefont {{You}}, \citenamefont {{Lu}},\ and\
  \citenamefont {{Xu}}}]{Liu_2017}%
  \BibitemOpen
  \bibfield  {author} {\bibinfo {author} {\bibfnamefont {M.}~\bibnamefont
  {{Liu}}}, \bibinfo {author} {\bibfnamefont {Z.}~\bibnamefont {{Zhao}}},
  \bibinfo {author} {\bibfnamefont {X.}~\bibnamefont {{You}}}, \bibinfo
  {author} {\bibfnamefont {J.}~\bibnamefont {{Lu}}}, \ and\ \bibinfo {author}
  {\bibfnamefont {L.}~\bibnamefont {{Xu}}},\ }\href {\doibase
  10.1016/j.physletb.2017.04.033} {\bibfield  {journal} {\bibinfo  {journal}
  {Physics Letters B}\ }\textbf {\bibinfo {volume} {770}},\ \bibinfo {pages}
  {8} (\bibinfo {year} {2017})}\BibitemShut {NoStop}%
\bibitem [{\citenamefont {{Wei}}\ \emph {et~al.}(2017)\citenamefont {{Wei}},
  \citenamefont {{Zhang}}, \citenamefont {{Wu}}, \citenamefont {{Gao}},
  \citenamefont {{M{\'e}sz{\'a}ros}}, \citenamefont {{Zhang}}, \citenamefont
  {{Dai}}, \citenamefont {{Zhang}},\ and\ \citenamefont {{Zhu}}}]{Wei_2017}%
  \BibitemOpen
  \bibfield  {author} {\bibinfo {author} {\bibfnamefont {J.-J.}\ \bibnamefont
  {{Wei}}}, \bibinfo {author} {\bibfnamefont {B.-B.}\ \bibnamefont {{Zhang}}},
  \bibinfo {author} {\bibfnamefont {X.-F.}\ \bibnamefont {{Wu}}}, \bibinfo
  {author} {\bibfnamefont {H.}~\bibnamefont {{Gao}}}, \bibinfo {author}
  {\bibfnamefont {P.}~\bibnamefont {{M{\'e}sz{\'a}ros}}}, \bibinfo {author}
  {\bibfnamefont {B.}~\bibnamefont {{Zhang}}}, \bibinfo {author} {\bibfnamefont
  {Z.-G.}\ \bibnamefont {{Dai}}}, \bibinfo {author} {\bibfnamefont {S.-N.}\
  \bibnamefont {{Zhang}}}, \ and\ \bibinfo {author} {\bibfnamefont {Z.-H.}\
  \bibnamefont {{Zhu}}},\ }\href {\doibase 10.1088/1475-7516/2017/11/035}
  {\bibfield  {journal} {\bibinfo  {journal} {\jcap}\ }\textbf {\bibinfo
  {volume} {2017}},\ \bibinfo {eid} {035} (\bibinfo {year} {2017})}\BibitemShut
  {NoStop}%
\bibitem [{\citenamefont {{Shoemaker}}\ and\ \citenamefont
  {{Murase}}(2018)}]{Shoemaker_2018}%
  \BibitemOpen
  \bibfield  {author} {\bibinfo {author} {\bibfnamefont {I.~M.}\ \bibnamefont
  {{Shoemaker}}}\ and\ \bibinfo {author} {\bibfnamefont {K.}~\bibnamefont
  {{Murase}}},\ }\href {\doibase 10.1103/PhysRevD.97.083013} {\bibfield
  {journal} {\bibinfo  {journal} {\prd}\ }\textbf {\bibinfo {volume} {97}},\
  \bibinfo {eid} {083013} (\bibinfo {year} {2018})}\BibitemShut {NoStop}%
\bibitem [{\citenamefont {Boran}\ \emph {et~al.}(2018)\citenamefont {Boran},
  \citenamefont {Desai}, \citenamefont {Kahya},\ and\ \citenamefont
  {Woodard}}]{Boran_2018}%
  \BibitemOpen
  \bibfield  {author} {\bibinfo {author} {\bibfnamefont {S.}~\bibnamefont
  {Boran}}, \bibinfo {author} {\bibfnamefont {S.}~\bibnamefont {Desai}},
  \bibinfo {author} {\bibfnamefont {E.~O.}\ \bibnamefont {Kahya}}, \ and\
  \bibinfo {author} {\bibfnamefont {R.~P.}\ \bibnamefont {Woodard}},\ }\href
  {\doibase 10.1103/PhysRevD.97.041501} {\bibfield  {journal} {\bibinfo
  {journal} {Phys. Rev. D}\ }\textbf {\bibinfo {volume} {97}},\ \bibinfo
  {pages} {041501} (\bibinfo {year} {2018})}\BibitemShut {NoStop}%
\bibitem [{\citenamefont {{Yao}}\ \emph {et~al.}(2020)\citenamefont {{Yao}},
  \citenamefont {{Zhao}}, \citenamefont {{Han}}, \citenamefont {{Wang}},
  \citenamefont {{Liu}},\ and\ \citenamefont {{Liu}}}]{Yao_2020}%
  \BibitemOpen
  \bibfield  {author} {\bibinfo {author} {\bibfnamefont {L.}~\bibnamefont
  {{Yao}}}, \bibinfo {author} {\bibfnamefont {Z.}~\bibnamefont {{Zhao}}},
  \bibinfo {author} {\bibfnamefont {Y.}~\bibnamefont {{Han}}}, \bibinfo
  {author} {\bibfnamefont {J.}~\bibnamefont {{Wang}}}, \bibinfo {author}
  {\bibfnamefont {T.}~\bibnamefont {{Liu}}}, \ and\ \bibinfo {author}
  {\bibfnamefont {M.}~\bibnamefont {{Liu}}},\ }\href {\doibase
  10.3847/1538-4357/abab02} {\bibfield  {journal} {\bibinfo  {journal} {\apj}\
  }\textbf {\bibinfo {volume} {900}},\ \bibinfo {eid} {31} (\bibinfo {year}
  {2020})}\BibitemShut {NoStop}%
\bibitem [{\citenamefont {{Yang}}\ \emph {et~al.}(2020)\citenamefont {{Yang}},
  \citenamefont {{Han}},\ and\ \citenamefont {{Wang}}}]{Yang_2020}%
  \BibitemOpen
  \bibfield  {author} {\bibinfo {author} {\bibfnamefont {S.-C.}\ \bibnamefont
  {{Yang}}}, \bibinfo {author} {\bibfnamefont {W.-B.}\ \bibnamefont {{Han}}}, \
  and\ \bibinfo {author} {\bibfnamefont {G.}~\bibnamefont {{Wang}}},\ }\href
  {\doibase 10.1093/mnrasl/slaa143} {\bibfield  {journal} {\bibinfo  {journal}
  {\mnras}\ }\textbf {\bibinfo {volume} {499}},\ \bibinfo {pages} {L53}
  (\bibinfo {year} {2020})}\BibitemShut {NoStop}%
\bibitem [{\citenamefont {Wang}\ \emph {et~al.}(2016)\citenamefont {Wang},
  \citenamefont {Liu},\ and\ \citenamefont {Wang}}]{Wang_2016}%
  \BibitemOpen
  \bibfield  {author} {\bibinfo {author} {\bibfnamefont {Z.-Y.}\ \bibnamefont
  {Wang}}, \bibinfo {author} {\bibfnamefont {R.-Y.}\ \bibnamefont {Liu}}, \
  and\ \bibinfo {author} {\bibfnamefont {X.-Y.}\ \bibnamefont {Wang}},\ }\href
  {\doibase 10.1103/PhysRevLett.116.151101} {\bibfield  {journal} {\bibinfo
  {journal} {Phys. Rev. Lett.}\ }\textbf {\bibinfo {volume} {116}},\ \bibinfo
  {pages} {151101} (\bibinfo {year} {2016})}\BibitemShut {NoStop}%
\bibitem [{\citenamefont {{Wei}}\ \emph {et~al.}(2016)\citenamefont {{Wei}},
  \citenamefont {{Wang}}, \citenamefont {{Gao}},\ and\ \citenamefont
  {{Wu}}}]{Wei_2016b}%
  \BibitemOpen
  \bibfield  {author} {\bibinfo {author} {\bibfnamefont {J.-J.}\ \bibnamefont
  {{Wei}}}, \bibinfo {author} {\bibfnamefont {J.-S.}\ \bibnamefont {{Wang}}},
  \bibinfo {author} {\bibfnamefont {H.}~\bibnamefont {{Gao}}}, \ and\ \bibinfo
  {author} {\bibfnamefont {X.-F.}\ \bibnamefont {{Wu}}},\ }\href {\doibase
  10.3847/2041-8205/818/1/L2} {\bibfield  {journal} {\bibinfo  {journal}
  {\apjl}\ }\textbf {\bibinfo {volume} {818}},\ \bibinfo {eid} {L2} (\bibinfo
  {year} {2016})}\BibitemShut {NoStop}%
\bibitem [{\citenamefont {{Boran}}\ \emph {et~al.}(2019)\citenamefont
  {{Boran}}, \citenamefont {{Desai}},\ and\ \citenamefont
  {{Kahya}}}]{Boran_2019}%
  \BibitemOpen
  \bibfield  {author} {\bibinfo {author} {\bibfnamefont {S.}~\bibnamefont
  {{Boran}}}, \bibinfo {author} {\bibfnamefont {S.}~\bibnamefont {{Desai}}}, \
  and\ \bibinfo {author} {\bibfnamefont {E.~O.}\ \bibnamefont {{Kahya}}},\
  }\href {\doibase 10.1140/epjc/s10052-019-6695-6} {\bibfield  {journal}
  {\bibinfo  {journal} {European Physical Journal C}\ }\textbf {\bibinfo
  {volume} {79}},\ \bibinfo {eid} {185} (\bibinfo {year} {2019})}\BibitemShut
  {NoStop}%
\bibitem [{\citenamefont {Laha}(2019)}]{Laha_2019}%
  \BibitemOpen
  \bibfield  {author} {\bibinfo {author} {\bibfnamefont {R.}~\bibnamefont
  {Laha}},\ }\href {\doibase 10.1103/PhysRevD.100.103002} {\bibfield  {journal}
  {\bibinfo  {journal} {Phys. Rev. D}\ }\textbf {\bibinfo {volume} {100}},\
  \bibinfo {pages} {103002} (\bibinfo {year} {2019})}\BibitemShut {NoStop}%
\bibitem [{\citenamefont {Wei}\ \emph {et~al.}(2019)\citenamefont {Wei},
  \citenamefont {Zhang}, \citenamefont {Shao}, \citenamefont {Gao},
  \citenamefont {Li}, \citenamefont {Yin}, \citenamefont {Wu}, \citenamefont
  {Wang}, \citenamefont {Zhang},\ and\ \citenamefont {Dai}}]{Wei_2019}%
  \BibitemOpen
  \bibfield  {author} {\bibinfo {author} {\bibfnamefont {J.-J.}\ \bibnamefont
  {Wei}}, \bibinfo {author} {\bibfnamefont {B.-B.}\ \bibnamefont {Zhang}},
  \bibinfo {author} {\bibfnamefont {L.}~\bibnamefont {Shao}}, \bibinfo {author}
  {\bibfnamefont {H.}~\bibnamefont {Gao}}, \bibinfo {author} {\bibfnamefont
  {Y.}~\bibnamefont {Li}}, \bibinfo {author} {\bibfnamefont {Q.-Q.}\
  \bibnamefont {Yin}}, \bibinfo {author} {\bibfnamefont {X.-F.}\ \bibnamefont
  {Wu}}, \bibinfo {author} {\bibfnamefont {X.-Y.}\ \bibnamefont {Wang}},
  \bibinfo {author} {\bibfnamefont {B.}~\bibnamefont {Zhang}}, \ and\ \bibinfo
  {author} {\bibfnamefont {Z.-G.}\ \bibnamefont {Dai}},\ }\href {\doibase
  https://doi.org/10.1016/j.jheap.2019.01.002} {\bibfield  {journal} {\bibinfo
  {journal} {Journal of High Energy Astrophysics}\ }\textbf {\bibinfo {volume}
  {22}},\ \bibinfo {pages} {1} (\bibinfo {year} {2019})}\BibitemShut {NoStop}%
\bibitem [{\citenamefont {{Yang}}\ and\ \citenamefont
  {{Zhang}}(2016)}]{Yang_2016}%
  \BibitemOpen
  \bibfield  {author} {\bibinfo {author} {\bibfnamefont {Y.-P.}\ \bibnamefont
  {{Yang}}}\ and\ \bibinfo {author} {\bibfnamefont {B.}~\bibnamefont
  {{Zhang}}},\ }\href {\doibase 10.1103/PhysRevD.94.101501} {\bibfield
  {journal} {\bibinfo  {journal} {\prd}\ }\textbf {\bibinfo {volume} {94}},\
  \bibinfo {eid} {101501} (\bibinfo {year} {2016})}\BibitemShut {NoStop}%
\bibitem [{\citenamefont {{Zhang}}\ and\ \citenamefont
  {{Gong}}(2017)}]{Zhang_2017}%
  \BibitemOpen
  \bibfield  {author} {\bibinfo {author} {\bibfnamefont {Y.}~\bibnamefont
  {{Zhang}}}\ and\ \bibinfo {author} {\bibfnamefont {B.}~\bibnamefont
  {{Gong}}},\ }\href {\doibase 10.3847/1538-4357/aa61fb} {\bibfield  {journal}
  {\bibinfo  {journal} {\apj}\ }\textbf {\bibinfo {volume} {837}},\ \bibinfo
  {eid} {134} (\bibinfo {year} {2017})}\BibitemShut {NoStop}%
\bibitem [{\citenamefont {{Desai}}\ and\ \citenamefont
  {{Kahya}}(2018)}]{Desai_2018}%
  \BibitemOpen
  \bibfield  {author} {\bibinfo {author} {\bibfnamefont {S.}~\bibnamefont
  {{Desai}}}\ and\ \bibinfo {author} {\bibfnamefont {E.}~\bibnamefont
  {{Kahya}}},\ }\href {\doibase 10.1140/epjc/s10052-018-5571-0} {\bibfield
  {journal} {\bibinfo  {journal} {European Physical Journal C}\ }\textbf
  {\bibinfo {volume} {78}},\ \bibinfo {eid} {86} (\bibinfo {year}
  {2018})}\BibitemShut {NoStop}%
\bibitem [{\citenamefont {{Leung}}\ \emph {et~al.}(2018)\citenamefont
  {{Leung}}, \citenamefont {{Hu}}, \citenamefont {{Harris}}, \citenamefont
  {{Brown}}, \citenamefont {{Gallicchio}},\ and\ \citenamefont
  {{Nguyen}}}]{Leung_2018}%
  \BibitemOpen
  \bibfield  {author} {\bibinfo {author} {\bibfnamefont {C.}~\bibnamefont
  {{Leung}}}, \bibinfo {author} {\bibfnamefont {B.}~\bibnamefont {{Hu}}},
  \bibinfo {author} {\bibfnamefont {S.}~\bibnamefont {{Harris}}}, \bibinfo
  {author} {\bibfnamefont {A.}~\bibnamefont {{Brown}}}, \bibinfo {author}
  {\bibfnamefont {J.}~\bibnamefont {{Gallicchio}}}, \ and\ \bibinfo {author}
  {\bibfnamefont {H.}~\bibnamefont {{Nguyen}}},\ }\href {\doibase
  10.3847/1538-4357/aac954} {\bibfield  {journal} {\bibinfo  {journal} {\apj}\
  }\textbf {\bibinfo {volume} {861}},\ \bibinfo {eid} {66} (\bibinfo {year}
  {2018})}\BibitemShut {NoStop}%
\bibitem [{\citenamefont {{Wei}}\ and\ \citenamefont {{Wu}}(2021)}]{Wei_2021}%
  \BibitemOpen
  \bibfield  {author} {\bibinfo {author} {\bibfnamefont {J.-J.}\ \bibnamefont
  {{Wei}}}\ and\ \bibinfo {author} {\bibfnamefont {X.-F.}\ \bibnamefont
  {{Wu}}},\ }\href {\doibase 10.1007/s11467-021-1049-x} {\bibfield  {journal}
  {\bibinfo  {journal} {Frontiers of Physics}\ }\textbf {\bibinfo {volume}
  {16}},\ \bibinfo {eid} {44300} (\bibinfo {year} {2021})}\BibitemShut
  {NoStop}%
\bibitem [{\citenamefont {{Wang}}\ \emph {et~al.}(2017)\citenamefont {{Wang}},
  \citenamefont {{Zhang}}, \citenamefont {{Wang}}, \citenamefont {{Shen}},
  \citenamefont {{Liang}}, \citenamefont {{Li}}, \citenamefont {{Liao}},
  \citenamefont {{Jin}}, \citenamefont {{Yuan}}, \citenamefont {{Zou}},
  \citenamefont {{Fan}},\ and\ \citenamefont {{Wei}}}]{Wang_2017}%
  \BibitemOpen
  \bibfield  {author} {\bibinfo {author} {\bibfnamefont {H.}~\bibnamefont
  {{Wang}}}, \bibinfo {author} {\bibfnamefont {F.-W.}\ \bibnamefont {{Zhang}}},
  \bibinfo {author} {\bibfnamefont {Y.-Z.}\ \bibnamefont {{Wang}}}, \bibinfo
  {author} {\bibfnamefont {Z.-Q.}\ \bibnamefont {{Shen}}}, \bibinfo {author}
  {\bibfnamefont {Y.-F.}\ \bibnamefont {{Liang}}}, \bibinfo {author}
  {\bibfnamefont {X.}~\bibnamefont {{Li}}}, \bibinfo {author} {\bibfnamefont
  {N.-H.}\ \bibnamefont {{Liao}}}, \bibinfo {author} {\bibfnamefont {Z.-P.}\
  \bibnamefont {{Jin}}}, \bibinfo {author} {\bibfnamefont {Q.}~\bibnamefont
  {{Yuan}}}, \bibinfo {author} {\bibfnamefont {Y.-C.}\ \bibnamefont {{Zou}}},
  \bibinfo {author} {\bibfnamefont {Y.-Z.}\ \bibnamefont {{Fan}}}, \ and\
  \bibinfo {author} {\bibfnamefont {D.-M.}\ \bibnamefont {{Wei}}},\ }\href
  {\doibase 10.3847/2041-8213/aa9e08} {\bibfield  {journal} {\bibinfo
  {journal} {\apjl}\ }\textbf {\bibinfo {volume} {851}},\ \bibinfo {eid} {L18}
  (\bibinfo {year} {2017})}\BibitemShut {NoStop}%
\bibitem [{\citenamefont {{Lambert}}\ and\ \citenamefont {{Le
  Poncin-Lafitte}}(2009)}]{Lambert_2009}%
  \BibitemOpen
  \bibfield  {author} {\bibinfo {author} {\bibfnamefont {S.~B.}\ \bibnamefont
  {{Lambert}}}\ and\ \bibinfo {author} {\bibfnamefont {C.}~\bibnamefont {{Le
  Poncin-Lafitte}}},\ }\href {\doibase 10.1051/0004-6361/200911714} {\bibfield
  {journal} {\bibinfo  {journal} {\aap}\ }\textbf {\bibinfo {volume} {499}},\
  \bibinfo {pages} {331} (\bibinfo {year} {2009})}\BibitemShut {NoStop}%
\bibitem [{\citenamefont {{Lambert}}\ and\ \citenamefont {{Le
  Poncin-Lafitte}}(2011)}]{Lambert_2011}%
  \BibitemOpen
  \bibfield  {author} {\bibinfo {author} {\bibfnamefont {S.~B.}\ \bibnamefont
  {{Lambert}}}\ and\ \bibinfo {author} {\bibfnamefont {C.}~\bibnamefont {{Le
  Poncin-Lafitte}}},\ }\href {\doibase 10.1051/0004-6361/201016370} {\bibfield
  {journal} {\bibinfo  {journal} {\aap}\ }\textbf {\bibinfo {volume} {529}},\
  \bibinfo {eid} {A70} (\bibinfo {year} {2011})}\BibitemShut {NoStop}%
\bibitem [{\citenamefont {{Laureijs}}\ \emph {et~al.}(2011)\citenamefont
  {{Laureijs}} \emph {et~al.}}]{Euclid_2011}%
  \BibitemOpen
  \bibfield  {author} {\bibinfo {author} {\bibfnamefont {R.}~\bibnamefont
  {{Laureijs}}} \emph {et~al.},\ }\href@noop {} {\bibfield  {journal} {\bibinfo
   {journal} {arXiv e-prints}\ ,\ \bibinfo {eid} {arXiv:1110.3193}} (\bibinfo
  {year} {2011})}\BibitemShut {NoStop}%
\bibitem [{\citenamefont {{LSST Science Collaboration}}(2009)}]{LSST_2009}%
  \BibitemOpen
  \bibfield  {author} {\bibinfo {author} {\bibnamefont {{LSST Science
  Collaboration}}},\ }\href@noop {} {\bibfield  {journal} {\bibinfo  {journal}
  {arXiv e-prints}\ ,\ \bibinfo {eid} {arXiv:0912.0201}} (\bibinfo {year}
  {2009})}\BibitemShut {NoStop}%
\bibitem [{\citenamefont {{Amelino-Camelia}}\ \emph {et~al.}(1998)\citenamefont
  {{Amelino-Camelia}}, \citenamefont {{Ellis}}, \citenamefont {{Mavromatos}},
  \citenamefont {{Nanopoulos}},\ and\ \citenamefont
  {{Sarkar}}}]{AmelinoCamelia_1998}%
  \BibitemOpen
  \bibfield  {author} {\bibinfo {author} {\bibfnamefont {G.}~\bibnamefont
  {{Amelino-Camelia}}}, \bibinfo {author} {\bibfnamefont {J.}~\bibnamefont
  {{Ellis}}}, \bibinfo {author} {\bibfnamefont {N.~E.}\ \bibnamefont
  {{Mavromatos}}}, \bibinfo {author} {\bibfnamefont {D.~V.}\ \bibnamefont
  {{Nanopoulos}}}, \ and\ \bibinfo {author} {\bibfnamefont {S.}~\bibnamefont
  {{Sarkar}}},\ }\href {\doibase 10.1038/31647} {\bibfield  {journal} {\bibinfo
   {journal} {\nat}\ }\textbf {\bibinfo {volume} {393}},\ \bibinfo {pages}
  {763} (\bibinfo {year} {1998})}\BibitemShut {NoStop}%
\bibitem [{\citenamefont {{Jacob}}\ and\ \citenamefont
  {{Piran}}(2008)}]{Jacob_2008}%
  \BibitemOpen
  \bibfield  {author} {\bibinfo {author} {\bibfnamefont {U.}~\bibnamefont
  {{Jacob}}}\ and\ \bibinfo {author} {\bibfnamefont {T.}~\bibnamefont
  {{Piran}}},\ }\href {\doibase 10.1088/1475-7516/2008/01/031} {\bibfield
  {journal} {\bibinfo  {journal} {\jcap}\ }\textbf {\bibinfo {volume} {2008}},\
  \bibinfo {eid} {031} (\bibinfo {year} {2008})}\BibitemShut {NoStop}%
\bibitem [{\citenamefont {{Giani}}\ and\ \citenamefont
  {{Frion}}(2020)}]{Leonardo_2020}%
  \BibitemOpen
  \bibfield  {author} {\bibinfo {author} {\bibfnamefont {L.}~\bibnamefont
  {{Giani}}}\ and\ \bibinfo {author} {\bibfnamefont {E.}~\bibnamefont
  {{Frion}}},\ }\href {\doibase 10.1088/1475-7516/2020/09/008} {\bibfield
  {journal} {\bibinfo  {journal} {\jcap}\ }\textbf {\bibinfo {volume} {2020}},\
  \bibinfo {eid} {008} (\bibinfo {year} {2020})}\BibitemShut {NoStop}%
\bibitem [{\citenamefont {{Vasileiou}}\ \emph {et~al.}(2013)\citenamefont
  {{Vasileiou}}, \citenamefont {{Jacholkowska}}, \citenamefont {{Piron}},
  \citenamefont {{Bolmont}}, \citenamefont {{Couturier}}, \citenamefont
  {{Granot}}, \citenamefont {{Stecker}}, \citenamefont {{Cohen-Tanugi}},\ and\
  \citenamefont {{Longo}}}]{Vasileiou_2013}%
  \BibitemOpen
  \bibfield  {author} {\bibinfo {author} {\bibfnamefont {V.}~\bibnamefont
  {{Vasileiou}}}, \bibinfo {author} {\bibfnamefont {A.}~\bibnamefont
  {{Jacholkowska}}}, \bibinfo {author} {\bibfnamefont {F.}~\bibnamefont
  {{Piron}}}, \bibinfo {author} {\bibfnamefont {J.}~\bibnamefont {{Bolmont}}},
  \bibinfo {author} {\bibfnamefont {C.}~\bibnamefont {{Couturier}}}, \bibinfo
  {author} {\bibfnamefont {J.}~\bibnamefont {{Granot}}}, \bibinfo {author}
  {\bibfnamefont {F.~W.}\ \bibnamefont {{Stecker}}}, \bibinfo {author}
  {\bibfnamefont {J.}~\bibnamefont {{Cohen-Tanugi}}}, \ and\ \bibinfo {author}
  {\bibfnamefont {F.}~\bibnamefont {{Longo}}},\ }\href {\doibase
  10.1103/PhysRevD.87.122001} {\bibfield  {journal} {\bibinfo  {journal}
  {\prd}\ }\textbf {\bibinfo {volume} {87}},\ \bibinfo {eid} {122001} (\bibinfo
  {year} {2013})}\BibitemShut {NoStop}%
\bibitem [{\citenamefont {{Rubin}}\ \emph {et~al.}(2020)\citenamefont
  {{Rubin}}, \citenamefont {{Szapudi}}, \citenamefont {{Shappee}},\ and\
  \citenamefont {{Anand}}}]{Rubin_2020}%
  \BibitemOpen
  \bibfield  {author} {\bibinfo {author} {\bibfnamefont {D.}~\bibnamefont
  {{Rubin}}}, \bibinfo {author} {\bibfnamefont {I.}~\bibnamefont {{Szapudi}}},
  \bibinfo {author} {\bibfnamefont {B.~J.}\ \bibnamefont {{Shappee}}}, \ and\
  \bibinfo {author} {\bibfnamefont {G.~S.}\ \bibnamefont {{Anand}}},\ }\href
  {\doibase 10.3847/2041-8213/ab7018} {\bibfield  {journal} {\bibinfo
  {journal} {\apjl}\ }\textbf {\bibinfo {volume} {890}},\ \bibinfo {eid} {L6}
  (\bibinfo {year} {2020})}\BibitemShut {NoStop}%
\bibitem [{\citenamefont {{Hu}}\ and\ \citenamefont
  {{Cooray}}(2000)}]{Hu_2000}%
  \BibitemOpen
  \bibfield  {author} {\bibinfo {author} {\bibfnamefont {W.}~\bibnamefont
  {{Hu}}}\ and\ \bibinfo {author} {\bibfnamefont {A.}~\bibnamefont
  {{Cooray}}},\ }\href {\doibase 10.1103/PhysRevD.63.023504} {\bibfield
  {journal} {\bibinfo  {journal} {\prd}\ }\textbf {\bibinfo {volume} {63}},\
  \bibinfo {eid} {023504} (\bibinfo {year} {2000})}\BibitemShut {NoStop}%
\bibitem [{\citenamefont {{Li}}\ \emph {et~al.}(2019)\citenamefont {{Li}},
  \citenamefont {{Dodelson}},\ and\ \citenamefont {{Hu}}}]{Li_2019}%
  \BibitemOpen
  \bibfield  {author} {\bibinfo {author} {\bibfnamefont {P.}~\bibnamefont
  {{Li}}}, \bibinfo {author} {\bibfnamefont {S.}~\bibnamefont {{Dodelson}}}, \
  and\ \bibinfo {author} {\bibfnamefont {W.}~\bibnamefont {{Hu}}},\ }\href
  {\doibase 10.1103/PhysRevD.100.043502} {\bibfield  {journal} {\bibinfo
  {journal} {\prd}\ }\textbf {\bibinfo {volume} {100}},\ \bibinfo {eid}
  {043502} (\bibinfo {year} {2019})}\BibitemShut {NoStop}%
\bibitem [{\citenamefont {Valentino}\ \emph {et~al.}(2020)\citenamefont
  {Valentino} \emph {et~al.}}]{Cosmology_Intertwined_1}%
  \BibitemOpen
  \bibfield  {author} {\bibinfo {author} {\bibfnamefont {E.~D.}\ \bibnamefont
  {Valentino}} \emph {et~al.},\ }\href@noop {} {} (\bibinfo {year} {2020}),\
  \Eprint {http://arxiv.org/abs/2008.11284} {arXiv:2008.11284} \BibitemShut
  {NoStop}%
\end{thebibliography}%

\end{document}